\documentclass[anual]{acmsiggraph}

\TOGonlineid{0257}

\TOGvolume{0}
\TOGnumber{0}

\title{GRASS: Generative Recursive Autoencoders for Shape Structures}

\author{
Jun Li$^{1}$\hspace{10pt}
Kai Xu$^{1,2,3}$\thanks{Corresponding author: kevin.kai.xu@gmail.com}\hspace{10pt}
Siddhartha Chaudhuri$^{4}$ \hspace{10pt}
Ersin Yumer$^{5}$ \hspace{10pt}
Hao Zhang$^{6}$ \hspace{10pt}
Leonidas Guibas$^{7}$ \hspace{10pt}
\vspace{4pt}\\
$^1$National University of Defense Technology\hspace{10pt}
$^2$Shenzhen University\hspace{10pt}
$^3$Shandong University\hspace{10pt}
$^4$IIT Bombay
\\
$^5$Adobe Research\hspace{10pt}
$^6$Simon Fraser University\hspace{10pt}
$^7$Stanford University
}

\pdfauthor{Jun Li, Kai Xu, Siddhartha Chaudhuri, Ersin Yumer, Hao Zhang, Leonidas Guibas}


\usepackage{times} 
\usepackage{mathptmx}
\usepackage{mathtools}
\usepackage{amsmath, amsthm, amsfonts, amssymb}
\usepackage{mathrsfs}
\usepackage{graphicx}
\usepackage{textcomp}
\usepackage{wrapfig}
\usepackage{subfig}
\usepackage{parskip}
\usepackage{helvet}
\usepackage{times}
\usepackage{color}
\usepackage{xspace}
\usepackage{overpic}
\usepackage{subfig}
\usepackage{wrapfig}
\usepackage{enumitem}

\definecolor{turquoise}{cmyk}{0.65,0,0.1,0.1}
\definecolor{purple}{rgb}{0.65,0,0.65}
\definecolor{dark_green}{rgb}{0, 0.5, 0}
\definecolor{orange}{rgb}{0.8, 0.6, 0.2}
\definecolor{red}{rgb}{0.8, 0.2, 0.2}
\definecolor{brown}{rgb}{0.5, 0.16, 0.16}
\newcommand{\kx}[1]{{\color{dark_green}#1}}

\newcommand{\ersin}[1]{{\color{cyan}#1}}
\newcommand{\rz}[1]{{\color{blue}#1}}

\newcommand{\leo}[1]{{\color{red}#1}}

\newcommand{\R}{\mathbb{R}}

\newcommand{\AdjEnc}{\mbox{\sc AdjEnc}\xspace}
\newcommand{\AdjDec}{\mbox{\sc AdjDec}\xspace}
\newcommand{\SymEnc}{\mbox{\sc SymEnc}\xspace}
\newcommand{\SymDec}{\mbox{\sc SymDec}\xspace}
\newcommand{\BoxEnc}{\mbox{\sc BoxEnc}\xspace}
\newcommand{\BoxDec}{\mbox{\sc BoxDec}\xspace}
\newcommand{\NodeCat}{\mbox{\sc NodeClsfr}\xspace}
\newcommand{\GeoEnc}{\mbox{\sc GeoEnc}\xspace}
\newcommand{\GeoDec}{\mbox{\sc GeoDec}\xspace}
\newcommand{\GeoMap}{\mbox{\sc GeoMap}\xspace}

\newcommand{\mypara}{\vspace*{-5pt}\paragraph}

\setlist[description]{leftmargin=3mm}
\setlist[itemize]{leftmargin=3mm}

\begin{document}



\teaser{
   \begin{overpic}[width=1.0\textwidth,tics=100]{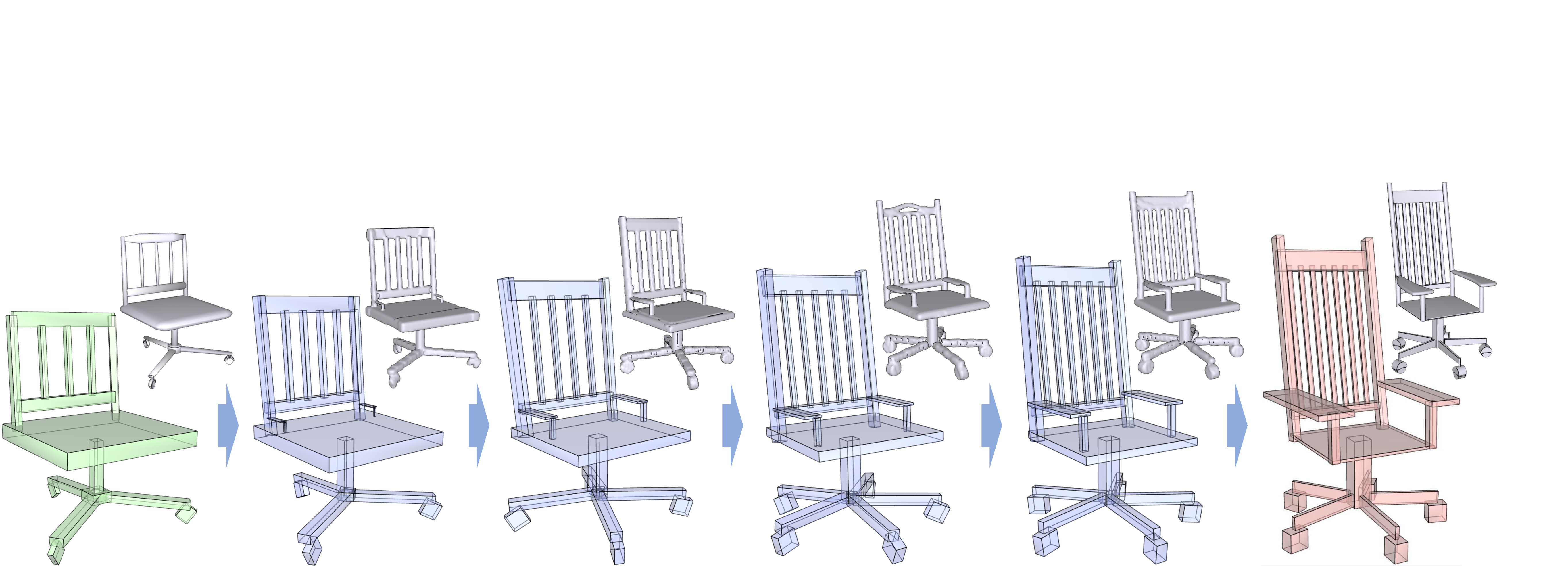}
   \end{overpic}
   \caption{We develop GRASS, a Generative Recursive Autoencoder for Shape Structures, which enables {\em structural\/} blending between two 3D shapes. Note the discrete
   blending of translational symmetries (slats on the chair backs) and rotational symmetries (the swivel legs). GRASS encodes and synthesizes box structures (bottom) and
   part geometries (top) separately. The blending is performed on fixed-length codes learned by the unsupervised autoencoder, without any form of part correspondences, given or computed.}
   \label{fig:teaser}
}

\maketitle


\begin{abstract}
We introduce a novel neural network architecture for {\em encoding\/} and {\em synthesis\/} of 3D shapes, particularly their {\em structures\/}. Our key insight is that 3D shapes are effectively characterized by their {\em hierarchical\/} organization of parts, which reflects fundamental intra-shape relationships such as adjacency and symmetry. We develop a {\em recursive\/} neural net (RvNN) based autoencoder to map a flat, unlabeled, arbitrary part layout to a compact code.
The code effectively captures hierarchical structures of man-made 3D objects of varying structural complexities despite being fixed-dimensional: an associated decoder maps a code back to a full hierarchy. The learned bidirectional mapping is further tuned using an adversarial setup to yield a generative model of plausible structures, from which novel structures can be sampled. Finally, our structure synthesis framework is augmented by a second trained module that produces fine-grained part geometry, conditioned on global and local structural context, leading to a full generative pipeline for 3D shapes. We demonstrate that without supervision, our network learns meaningful structural hierarchies adhering to perceptual grouping principles, produces compact codes which enable applications
such as shape classification and partial matching, and supports shape synthesis and interpolation with significant variations in topology and geometry.


\end{abstract}




\keywordlist




\section{Introduction}
\label{sec:intro}

\begin{figure*}[t] \centering
	\begin{overpic}[width=1.0\textwidth,tics=100]{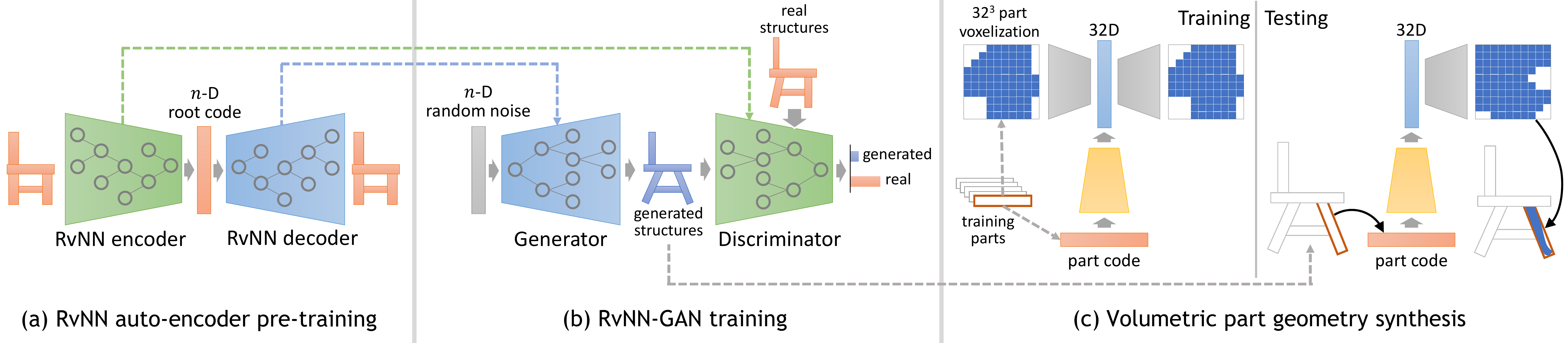}
	\end{overpic}
    \caption{An overview of our pipeline, including the three key stages: (a) pre-training the RvNN autoencoder to obtain root codes for shapes, (b) using a GAN module to learn the actual shape manifold within the code space, and (c) using a second network to convert synthesized OBBs to detailed geometry.
    }
    \label{fig:overview}
\end{figure*}

Recent progress on training neural networks for image~\cite{oord2016_rnn} and
speech~\cite{oord2016_wavenet} synthesis has led many to ask whether a
similar success is achievable in learning generative models for 3D shapes.
While an image is most naturally viewed as a 2D signal of pixel values and a
piece of speech as a sampled 1D audio wave, the question of what is {\em the\/} canonical
representation for 3D shapes (voxels, surfaces meshes, or multi-view images)
may not always yield a consensus answer. Unlike images or sound, a 3D shape does
not have a natural parameterization over a regular low-dimensional grid. Further, many
3D shapes, especially of man-made artifacts, are highly structured (e.g. with
hierarchical decompositions and nested symmetries), while exhibiting rich structural variations
even within the same object class (e.g. consider the variety of chairs). Hence, the
stationarity and compositionality assumptions~\cite{henaff2015} behind the success
of most neural nets for {\em natural\/} images or speech are no longer applicable.

In this paper, we are interested in learning {\em generative neural nets\/} for
{\em structured\/} shape representations of man-made 3D objects. In general,
shape structures are defined by the arrangement of, and relations between, shape
parts~\cite{mitra2013}. Developing neural nets for structured shape representations
requires a significant departure from existing works on convolutional neural networks
(CNNs) for volumetric~\cite{wu2015,girdhar2016,yumer2016,wu2016} or
view-based~\cite{su2015,qi2016,sinha2016} shape representations. These works
primarily adapt classical CNN architectures for image analysis. They do not explicitly encode or
synthesize part arrangements or relations such as symmetries.

Our goal is to learn a generative neural net for shape structures characterizing
an object class, e.g. chairs or candelabras. The main
challenges we face are two-fold. The first is how to properly ``mix'', or
{\em jointly\/} encode and synthesize (discrete) structure and (continuous) geometry.
The second is due to intra-class structural variations. If we treat
shape structures as graphs, the foremost question is how to enable a generic
neural network to work with graphs of {\em different\/} combinatorial structures and sizes. Both challenges are
unique to our problem setting and neither has been addressed by networks which take inputs in
the form of unstructured, fixed-size, low-dimensional grid data, e.g. images or volumes.

Our key insight is that most shape structures are naturally {\em hierarchical\/} and
hierarchies can jointly encode structure and geometry. Most importantly, regardless of
the variations across shape structures, a coding scheme that recursively contracts hierarchy or tree
nodes into their parents attains {\em unification\/} at the top --- any finite set of structures
eventually collapses to root node codes with a possibly large but {\em fixed length\/}. We learn a neural
network which can recursively encode hierarchies into root codes and invert the
process via decoding. Then, by further learning a distribution over the root codes for a class
of shapes, new root codes can be generated and decoded to synthesize new structures and shapes in that class.

Specifically, we represent a 3D shape using a {\em symmetry hierarchy\/}~\cite{wang2011},
which defines how parts in the shape are recursively grouped by symmetry and
assembled by connectivity. Our neural net architecture, which learns to infer such a hierarchy for a
shape in an unsupervised fashion, is inspired by the {\em recursive neural nets\/} (RvNN)\footnote{Note that we are adding the letter `v' to
the acronym RNN, since by now, the term RNN most frequently refers to {\em recurrent\/}
neural networks.} of Socher et al.~\shortcite{socher2011,socher2012} developed
for text and image understanding. By treating text as a set of words and an image as a set of
superpixels, an RvNN learns a parse tree which recursively merges text/image segments.
There are two key differences and challenges that come with our work:
\begin{itemize}
\item First, the RvNNs of Socher et al.~\shortcite{socher2011} always merge two
adjacent elements and this is modeled using the same network at every tree node.
However, in a symmetry hierarchy, grouping by symmetry and
assembly by connectivity are characteristically different merging operations.
As well, the network structures at a tree node must accommodate assembly,
reflectional symmetry, and rotational/translational symmetries of varying orders.
\item Second, our main goal is to learn a {\em generative\/} RvNN, for part-based
shape structures that are explicitly represented as discrete structural combinations
of geometric entities.
\end{itemize}
To accomplish these goals, we focus on learning an abstraction of symmetry
hierarchies, which are composed of spatial arrangements of oriented bounding
boxes (OBBs). Each OBB is defined by a fixed-length code to represent its
geometry and these codes sit at the leaves of the hierarchies. Internal nodes of the
hierarchies, also characterized by fixed-length codes, encode both the geometry
of its child OBBs and their detailed grouping mechanism: whether by connectivity or symmetry.

We pre-train an {\em unsupervised\/} RvNN using OBB arrangements endowed with
box connectivity and various types of symmetry.
%
Our neural network is an {\em autoencoder\/}-based RvNN which recursively assembles
or (symmetrically) groups a set of OBBs into a fixed-length root code and then
decodes the root to reconstruct the input; see Figure~\ref{fig:overview}(a).
The network comprises two types of nodes: one to handle assembly of connected parts,
and one to handle symmetry grouping. Each merging operation takes two or more OBBs
as input. Our RvNN learns how to best organize a shape structure into a symmetry
hierarchy to arrive at a compact and minimal-loss code accounting for
both geometry and structure.

To synthesize new 3D shapes, we extend the pre-trained autoencoder RvNN into
a generative model. We learn a distribution over root codes constructed
from shape structures for 3D objects of the same class, e.g. chairs. This
step utilizes a generative adversarial network (GAN), similar to a VAE-GAN~\cite{larsen2015}, to learn a
low-dimensional manifold of root codes; see
Figure~\ref{fig:overview}(b). We sample and then project a root code onto the manifold
to synthesize an OBB arrangement. In the final stage, the boxes are filled with part
geometries by another generative model which learns a mapping between box features
and voxel grids; see Figure~\ref{fig:overview}(c).
We refer to our overall generative neural network as a {\em generative recursive
autoencoder\/} for shape structures, or GRASS. Figure~\ref{fig:overview} provides
an overview of the complete architecture.

The main contributions of our work can be summarized as follows. 
\begin{itemize}
\item The first generative neural network model for structured 3D shape representations
--- GRASS. This is realized by an autoencoder RvNN which learns to encode and decode
shape structures via discovered symmetry hierarchies, followed by two generative models
trained to synthesize box-level symmetry hierarchies and volumetric part geometries,
respectively.
\item A novel RvNN architecture which extends the original RvNN of Socher et
al.~\shortcite{socher2011} by making it generative and capable of encoding a variety of merging
operations (i.e. assembly by connectivity and symmetry groupings of different types).
\item An {\em unsupervised\/} autoencoder RvNN which jointly learns and encodes the structure
and geometry of box layouts of varying sizes into {\em fixed-length\/} vectors.
\end{itemize}

We demonstrate that our network learns meaningful structural hierarchies
adhering to perceptual grouping principles, produces compact codes which enable applications
such as shape classification and partial matching, and supports generative models which lead
to shape synthesis and interpolation with significant variations in topology and geometry.



\section{Related work}
\label{sec:related}


Our work is related to prior works on statistical models of 3D shape structures, including recent works on applying deep neural networks to shape representation. These models can be disciminative or generative, and capture continuous or discrete variations. We review the most relevant works below. Since our focus is shape synthesis, we emphasize generative models in our discussion.

\mypara{Statistical shape representations.} Early works on capturing statistical variations of the human body explored smooth deformations of a fixed template \cite{Blanz1999,Allen2003,Anguelov2005}. Later papers addressed discrete variations at the part level, employing stochastic shape grammars coded by hand \cite{Mueller2006}, learned from a single training example \cite{Bokeloh2010}, or learned from multiple training examples \cite{Talton2012}. Parallel works explored the use of part-based Bayesian networks \cite{Chaudhuri2011,Kalogerakis2012} and modular templates \cite{kim2013,Fish2014} to represent both continuous and discrete variations. However, these methods are severely limited in the variety and complexity of part layouts they can generate, and typically only work well for shape families with a few consistently appearing parts and a restricted number of possible layouts.
In a different approach, Talton et al.~\shortcite{Talton2009} learn a probability distribution over a shape space generated by a procedure operating on a fixed set of parameters. We are also inspired by some non-statistical shape representations such as the work of Wang et al.~\shortcite{wang2011} and van Kaick et al.~\shortcite{vanKaick2013} on extracting hierarchical structure from a shape: our goal in this paper is to learn consistent, probabilistic, hierarchical representations automatically from unlabeled datasets. Mitra et al.~\shortcite{mitra2013} provide an overview of a range of further works on statistical and structure-aware shape representations.

\mypara{Deep models of 3D shapes.} Recently, the success of deep neural networks in computer vision, speech recognition, and natural language processing has inspired researchers to apply such models to 3D shape analysis. While these are of course statistical shape representations, their immediate relevance to this paper merits a separate section from the above. Most of these works have focused on extending computer vision techniques developed for images -- 2D grids of pixels -- to 3D grids of voxels. Wu et al.~\shortcite{wu2015} propose a generative model based on a deep belief network trained on a large, unannotated database of voxelized 3D shapes. They show applications of the model to shape synthesis and probabilistic shape completion for next-best view prediction. Girdhar et al.~\shortcite{girdhar2016} jointly train a deep convolutional encoder for 2D images and a deep convolutional decoder for voxelized 3D shapes, chained together so that the vector output of the encoder serves as the input code for the decoder, allowing 3D reconstruction from a 2D image. Yan et al.~\shortcite{Yan2016} propose a different encoder-decoder network for a similar application. Yumer and Mitra~\shortcite{yumer2016} present a 3D convolutional network that maps a voxelized shape plus a semantic modification intent to the deformation field required to realize that intent.

In a departure from voxel grids, Su et al.~\shortcite{su2015} build a powerful shape classifier based on multiple projected views of the object, by fine-tuning standard image-based CNNs trained on huge 2D datasets and applying a novel pooling mechanism. Masci et al.~\shortcite{Masci2015} build a convolutional network directly on non-Euclidean shape surfaces. Qi et al.~\shortcite{qi2016} discuss ways to improve the performance of both volumetric and multi-view CNNs for shape classification. In a recent work, Tulsiani et al.~\shortcite{tulsiani2017} develop a discriminative, CNN-based approach to consistently parse shapes into a bounded number of volumetric primitives.

We are inspired by the work of Huang et al.~\shortcite{huang2015}, who develop a deep Boltzmann machine-based model of 3D shape surfaces. This approach can be considered a spiritual successor of Kalogerakis et al.~\shortcite{Kalogerakis2012} and Kim et al.~\shortcite{kim2013}, learning modular templates that incorporate fine-grained part-level deformation models. In addition to being fully generative -- the model can be sampled for a point set representing an entirely new shape -- the method automatically refines shape correspondences and part boundaries during training. However, like the prior works, this approach is limited in the variety of layouts it can represent.

Wu et al.~\shortcite{wu2016} exploit the success of generative adversarial nets (GAN)~\cite{goodfellow2014} to improve upon the model of Wu et al.~\shortcite{wu2015}. At its core, their model is a generative decoder that takes as input a 200-D shape code and produces a voxel grid as output. The decoder is trained adversarially, and may be chained with a prior encoder that maps, say, a 2D image to the corresponding shape code. The method supports simple arithmetic and interpolation on the codes, enabling, for instance, topology-varying morphs between different shapes. Our work is complementary to this method: we seek to develop a powerful model of part layout variations that can accurately synthesize complex hierarchical structures beyond the representational power of low-resolution grids, can be trained on relatively fewer shapes, and is independent of voxel resolution.

\mypara{Neural models of graph structure.} The layout of parts of a shape inherently induces a non-Euclidean, graph-based topology defined by adjacency and relative placement. Several works, not concerning geometric analysis, have explored neural networks operating on graph domains. The most common such domains are of course linear chains defining text and speech signals. For these domains, recurrent neural networks (RNNs), as well as convolutional neural networks (CNNs) over sliding temporal windows, have proved very successful. Such linear models have even been adapted to generate non-linear output such as images, as in the work of van den Oord et al.~\shortcite{oord2016_rnn,oord2016_cnn}, producing the image row by row, pixel by pixel. These models are, however, limited in such adaptations since it is difficult to learn and enforce high-level graph-based organizational structure. Henaff et al.~\shortcite{henaff2015}, Duvenaud et al.~\shortcite{duvenaud2015} and Niepert et al.~\shortcite{niepert2016} propose convolutional networks that operate directly on arbitrary graphs by defining convolution as an operation on the radial neighborhood of a vertex. However, none of these works enable generative models. A different approach to this problem, which directly inspires our work, is the \emph{recursive} neural network (RvNN) proposed by Socher et al.~\shortcite{socher2011,socher2012}, which sequentially collapses edges of a graph to yield a hierarchy. We build upon the autoencoder version of this network, adapting it to learn the particular organizational principles that characterize 3D shape structure, and to extend it from a deterministic model to a probabilistic generative one.

\section{Overview}
\label{sec:overview}



Our method for learning GRASS, a hierarchical, symmetry-aware, generative model for 3D shapes, has three
stages, shown in Figure~\ref{fig:overview}. In this section, we summarize the stages and highlight important components and properties of the neural networks we use.

\mypara{Geometry and structure encoding.} We define an abstraction of symmetry
hierarchies, which are composed of spatial arrangements of oriented bounding
boxes (OBBs). Each OBB is defined by a fixed-length code to represent its
geometry. The fixed length code encodes both the geometry of its child OBBs
and their detailed grouping mechanism: whether by connectivity or symmetry.


\mypara{Stage 1: Recursive autoencoder.} In the first stage, we train an autoencoder for layouts of OBBs. The autoencoder maps a box layout with an arbitrary number and arrangement of components to a fixed-length root code that implicitly captures its salient features. The encoding is accomplished via a recursive neural network (RvNN) that repeatedly, in a bottom-up fashion, collapses a pair of boxes represented as codes into a merged code. The process also yields a hierarchical organizational structure for the boxes. The final code representing the entire layout is decoded to recover the boxes (plus the entire hierarchy) by an inverse process, and the training loss is measured in terms of a reconstruction error and back-propagated to update the network weights.


\mypara{Stage 2: Learning manifold of plausible structures.} We extend the autoencoder to a generative model of structures by learning a distribution over root codes that describes the shape manifold, or shape space, occupied by codes corresponding to meaningful shapes within the full code space. We train a generative adversarial model (GAN) for a low-dimensional manifold of root codes that can be decoded to structures indistinguishable, to an adversarial classifier, from the training set. Given a randomly selected root code, we project it to the GAN manifold to synthesize a plausible new structure.


\mypara{Stage 3: Part geometry synthesis.} In the final stage, the synthesized boxes are converted to actual shape parts. Given a box in a synthesized layout, we compute structure-aware recursive features that represent it in context. Then, we simultaneously learn a compact, invertible encoding of voxel grids representing part geometries as well as a mapping from contextual part features to the encoded voxelized geometry. This yields a procedure that can synthesize detailed geometry for a box in a shape structure.

By chaining together hierarchical structure generation and part geometry synthesis, we obtain the full GRASS pipeline for recursive synthesis of shape structures.

\section{Recursive model of shape structure}
\label{sec:rvnn}

In this section, we describe a method to encode shape structures into a short, fixed-dimensional code. The learned encoding is fully invertible, allowing the structure to be reconstructed from the code. In Section \ref{sec:gan}, we present our method to adversarially tune this structure decoder to map random codes to structures likely to come from real shapes. By combining this generator for sampling plausible shape structures with a method for synthesizing the geometry of individual parts (Section \ref{sec:partgeom}), we obtain our probabilistic generative model for 3D shapes.

Our key observation is that shape components are commonly arranged, or perceived to be arranged, hierarchically. This is a natural organizational principle in well-accepted theories of human cognition and design, which has been extensively leveraged computationally~\cite{Serre2013}. Perceptual and functional hierarchies follow patterns of component proximity and symmetry. Hence, the primary goal of our structural code is to successfully encode the hierarchical organization of the shape in terms of symmetries and adjacencies. An important metric of success is that the hierarchies are {\em consistent} across different shapes of the same category. We achieve this via a compact model of recursive component aggregation that tries to consistently identify similar substructures.

Our model is based on Recursive Autoencoders (RAE) for unlabeled binary trees, developed by Socher et al.~\shortcite{Socher2014}. The RAE framework proposed by Socher et al. consists of an {\em encoder} neural network that takes two $n$-dimensional inputs and produces a single $n$-dimensional output, and a {\em decoder} network that recovers two \mbox{$n$-D} vectors from a single \mbox{$n$-D} vector. In our experiments, $n = 80$.

Given a binary tree with \mbox{$n$-D} descriptors for the leaves, the RAE is used to recursively compute descriptors for the internal nodes, ending with a root code. The root code can be inverted to recover the original tree using the decoder, and a training loss formulated in terms of a reconstruction error for the leaves.

RAEs were originally intended for parsing natural language sentences in a discriminative setting, trained on unlabeled parse trees. We adapt this framework for the task of learning and synthesizing hierarchical shape structures. This requires several important technical contributions, including extending the framework to accommodate multiple encoder and decoder types, handling non-binary symmetric groups of parts, and probabilistically generating shapes (as described in the subsequent sections).

\begin{figure}[t] \centering
	\begin{overpic}[width=1.0\columnwidth,tics=100]{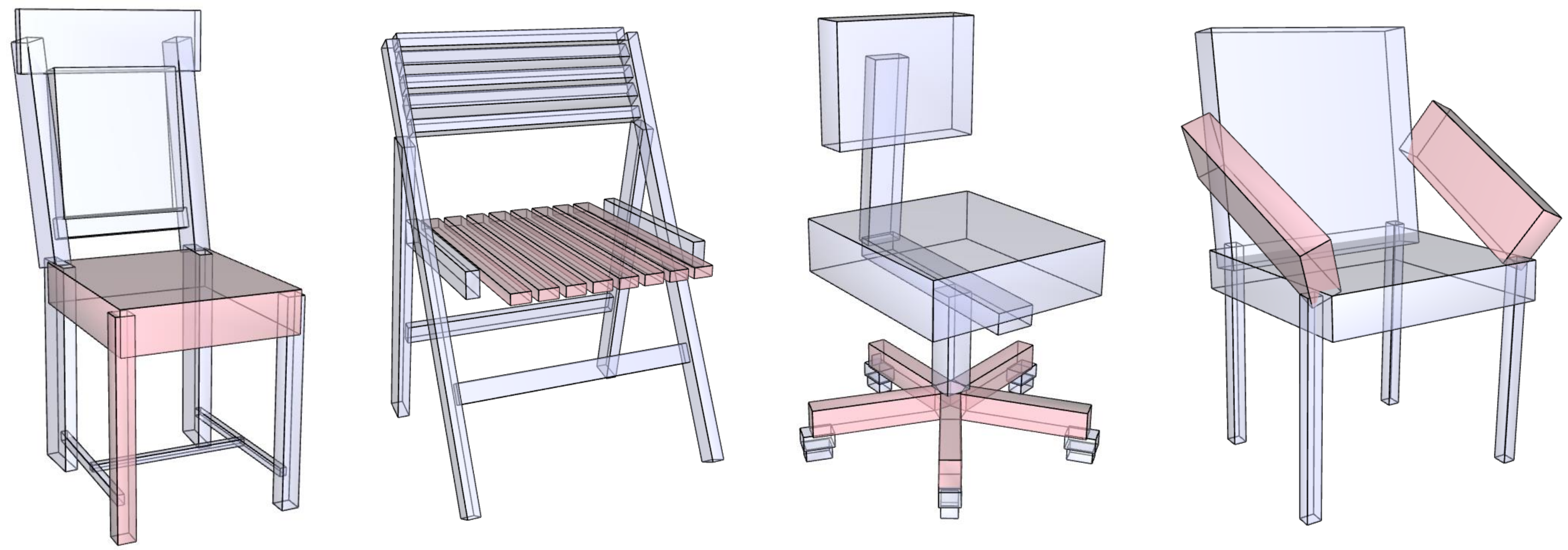}
	\end{overpic}
    \caption{Merging criteria used by our model demonstrated with 3D shapes represented by part bounding boxes (relevant parts highlighted in red). From left: (a) two adjacent parts, (b) translational symmetry, (c) rotational symmetry, and (d) reflective symmetry.}
    \label{fig:merges}
\end{figure}

\mypara{Criteria for recursive merging.} Our model of hierarchical organization of shape parts follows two common perceptual/cognitive criteria for recursive merging: a mergeable subset of parts is either an adjacent pair (the {\em adjacency} criterion) or part of a symmetry group (the {\em symmetry} criterion)\footnote{Currently, we make the reasonable single-object assumption that all parts are connected by either adjacency or symmetry. For disconnected, asymmetric shapes, we would need further merging criteria.}. An adjacent pair is represented by the bounding boxes of constituent parts. In this stage, we are interested only in representing the gross layout of parts, so we discard fine-grained geometric information and store only oriented part bounding boxes, following earlier work on shape layouts~\cite{Ovsjanikov11} --- fine-grained geometry synthesis is described in Section \ref{sec:partgeom}. We recognize three different types of symmetries, each represented by the bounding box of a generator part plus further parameters: (1) pairwise reflectional symmetry, parametrized by the plane of reflection; (2) $k$-fold rotational symmetry, parametrized by the number of parts $k$ and the axis of rotation; and (3) $k$-fold translational symmetry, parametrized by $k$ and the translation offset between parts. The different scenarios are illustrated in Figure \ref{fig:merges}. We generate training hierarchies that respect these criteria, and our autoencoder learns to synthesize hierarchies that follow them.

\mypara{Synthesizing training data.} To train our recursive autoencoder, we synthesize a large number of training hierarchies from a dataset of shapes. These shapes are assumed to be pre-segmented into constituent (unlabeled) parts, but do not have ground truth hierarchies. We adopt an iterative, randomized strategy to generate plausible hierarchies for a shape that satisfy the merging criteria described above. In each iteration, two or more parts are merged into a single one. A mergeable subset of parts is either adjacent or symmetric. We randomly sample a pair that satisfies one of the two criteria until no further merges are possible. In our experiments, we generated $20$ training hierarchies for each shape in this fashion. Note that none of these hierarchies is intended to represent ``ground truth''. Rather, they sample the space of plausible part groupings in a relatively unbiased fashion for training purposes.

\mypara{Autoencoder model.} To handle both adjacency and symmetry relations, our recursive autoencoder comprises two distinct types of encoder/decoder pairs. These types are:
\begin{description}
  \item[Adjacency.] The encoder for the adjacency module is a neural network \AdjEnc which merges codes for two adjacent parts into the code for a single part. It has two \mbox{$n$-D} inputs and one \mbox{$n$-D} output. Its parameters are a weight matrix $W_{ae} \in \R^{n \times 2n}$ and a bias vector $b_{ae} \in \R^n$, which are used to obtain the code of parent (merged) node $y$ from children $x_1$ and $x_2$ using the formula
  \[
    y \ = \ \tanh(W_{ae} \cdot [x_1 \ x_2] + b_{ae})
  \]
  The corresponding decoder \AdjDec splits a parent code $y$ back to child codes $x'_1$ and $x'_2$, using the reverse mapping
  \[
    [x'_1 \ x'_2] \ = \ \tanh(W_{ad} \cdot y + b_{ad})
  \]
  where $W_{ad} \in \R^{2n \times n}$ and $b_{ad} \in \R^{2n}$.

  \item[Symmetry.] The encoder for the symmetry module is a neural network \SymEnc which merges the $n$-D code for a generator part of a symmetry group, as well as the $m$-D parameters of the symmetry itself
      into a single $n$-D output. The code for a group with generator $x$ and parameters $p$ is computed as
  \[
    y \ = \ \tanh(W_{se} \cdot [x \ p] + b_{se})
  \]
  and the corresponding decoder \SymDec recovers the generator and symmetry parameters as
  \[
    [x' p'] \ = \ \tanh(W_{sd} \cdot y + b_{sd})
  \]
  where $W_{se} \in \R^{n \times (n + m)}$, $W_{sd} \in \R^{(n + m) \times n}$, $b_{se} \in \R^n$, and $b_{sd} \in \R^{m + n}$.
  In our implementation, we use $m = 8$ to encode symmetry parameters comprising symmetry type ($1$D); number of repetitions for rotational and translational symmetries ($1$D);
  and the mirror plane for reflective symmetry, rotation axis for rotational symmetry, or position and displacement for translational symmetry ($6$D).
\end{description}

In practice, the encoders/decoders for both adjacency and symmetry are implemented as two-layer networks, where the dimensions of the hidden and output layers are $100$D and $80$D, respectively.

The input to the recursive merging process is a collection of part bounding boxes. These need to be mapped to \mbox{$n$-D} vectors before they can be processed by the autoencoder. To this end, we employ additional single-layer neural networks \BoxEnc, which maps the $12$D parameters of a box (concatenating box center, dimensions and two axes) to an \mbox{$n$-D} code, and \BoxDec, which recovers the $12$D parameters from the \mbox{$n$-D} code. These networks are non-recursive, used simply to translate the input to the internal code representation at the beginning, and back again at the end.

Lastly, we jointly train an auxiliary classifier \NodeCat to decide which module to apply at each recursive decoding step. This classifier is a neural network with one hidden layer that takes as input the code of a node in the hierarchy, and outputs whether the node represents an adjacent pair of parts, a symmetry group, or a leaf node. Depending on the output of the classifier, either \AdjDec, \SymDec or \BoxDec is invoked.

\begin{figure}[t] \centering
	\begin{overpic}[width=1.0\columnwidth,tics=100]{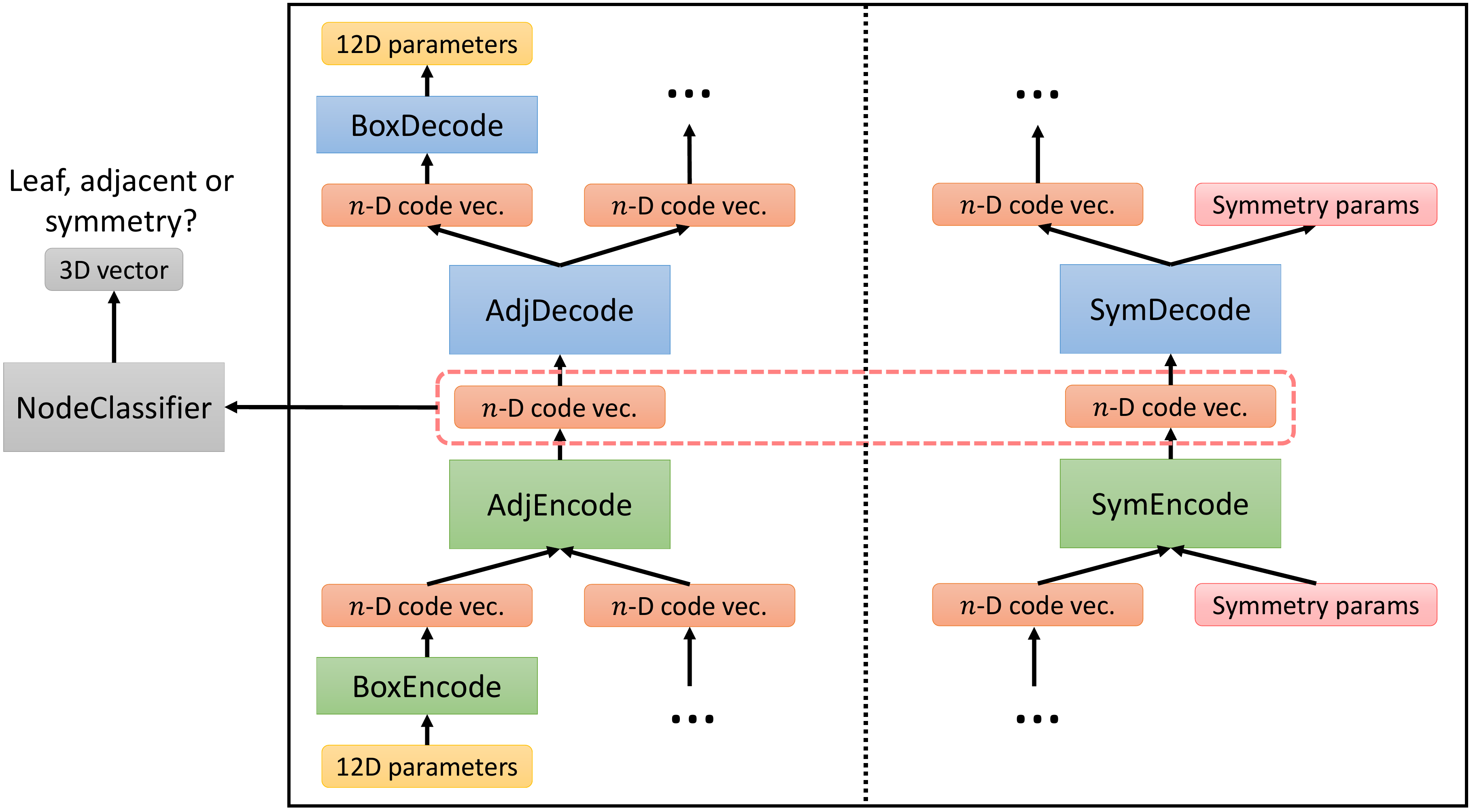}
	\end{overpic}
    \caption{Autoencoder training setup. Ellipsis dots indicate that the code could be either the output of \BoxEnc, \AdjEnc or \SymEnc, or the input to \BoxDec, \AdjDec or \SymDec.}
    \label{fig:ae_train}
\end{figure}

\begin{figure}[t] \centering
	\begin{overpic}[width=0.9\columnwidth,tics=100]{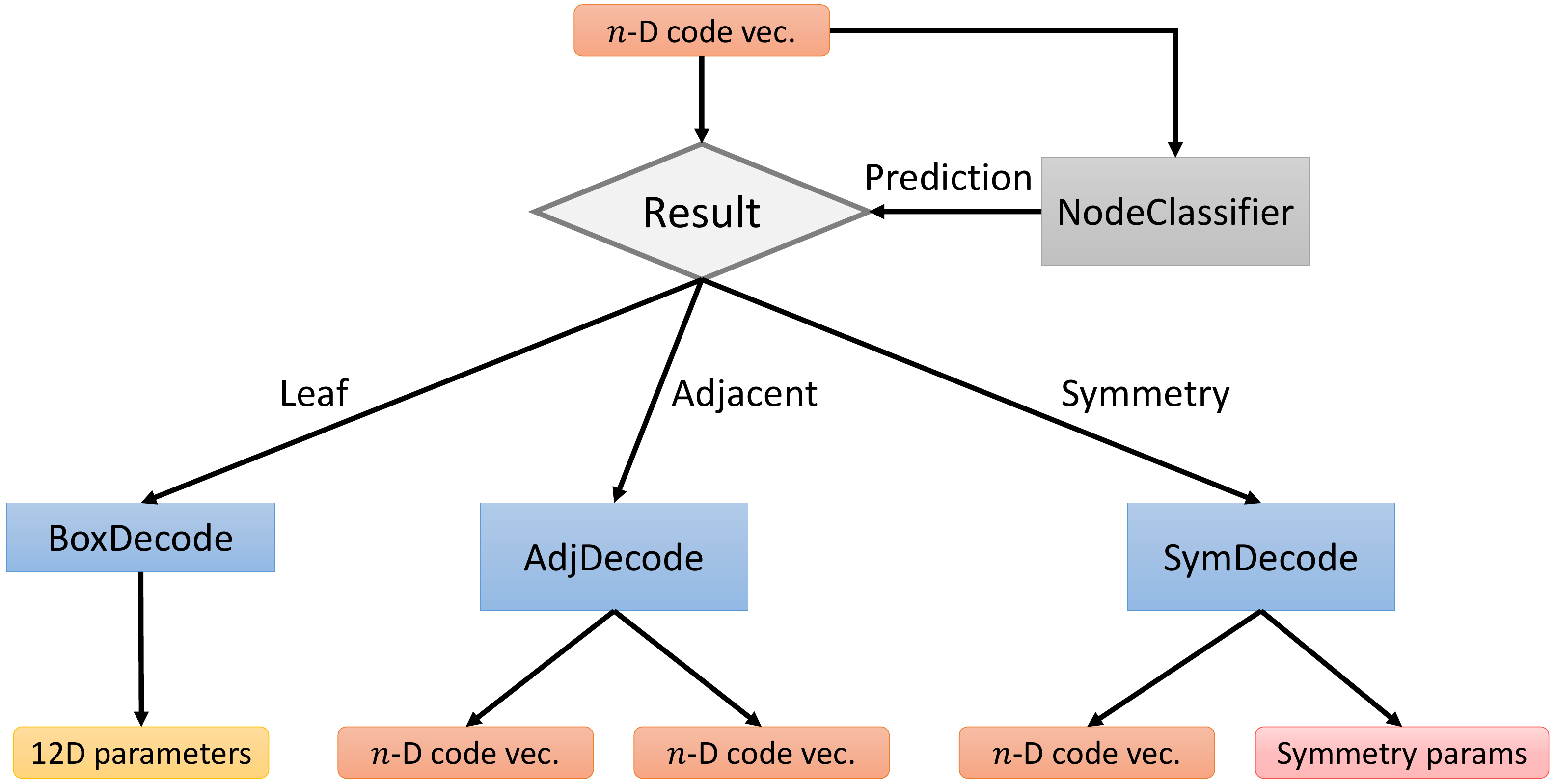}
	\end{overpic}
    \caption{Autoencoder test decoding setup.}
    \label{fig:ae_test}
\end{figure}

\mypara{Training.} To train our recursive autoencoder, we use BFGS with back-propagation, starting with a random initialization of weights sampled from a Gaussian distribution. The loss is formulated as a reconstruction error. Given a training hierarchy, we first encode each leaf-level part bounding box using \BoxEnc. Next, we recursively apply the corresponding encoder (\AdjEnc or \SymEnc) at each internal node until we obtain the code for the root. Finally, we invert the process, starting from the root code, to recover the leaf parameters by recursively applying the decoders \AdjDec and \SymDec, followed by a final application of \BoxDec. The loss is formulated as the sum of squared differences between the input and output parameters for each leaf box.

Note that during training (but not during testing), we use the input hierarchy for decoding, and hence always know which decoder to apply at which unfolded node, and the mapping between input and output boxes. We simultaneously train \NodeCat, with a three-class softmax classification with cross entropy loss, to recover the tree topology during testing. The training setup is illustrated in Figure \ref{fig:ae_train}.

\mypara{Testing.} During testing, we must address two distinct challenges. The first is to infer a plausible encoding hierarchy for a novel segmented shape without hierarchical organization. The second is to decode a given root code to recover the constituent bounding boxes of the shape.

\begin{figure}[b] \centering
	\begin{overpic}[width=1.0\columnwidth,tics=100]{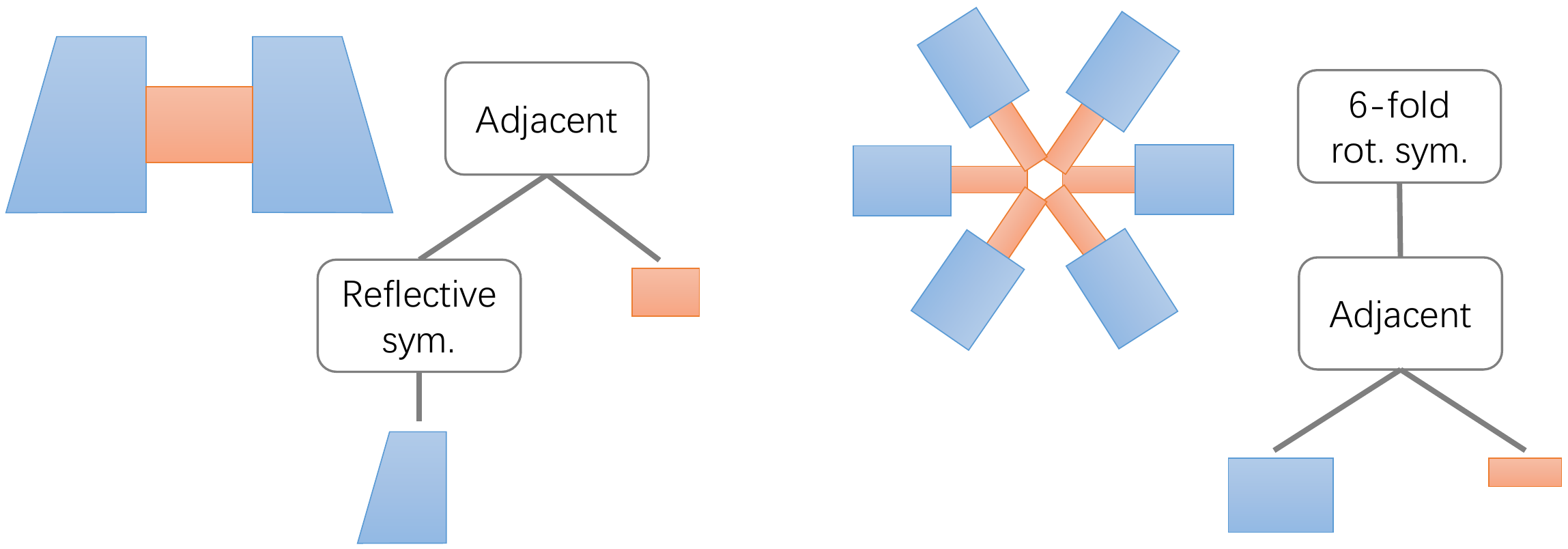}
	\end{overpic}
    \caption{Different two-step encoding orders for two examples, found by minimizing reconstruction errors during testing.
    Left: Symmetry (reflective) before adjacency. Right: Adjacency before symmetry (6-fold rotational).}
    \label{fig:encode_order}
\end{figure}

To infer a plausible hierarchy using the trained encoding modules, we resort to {\em greedy local search}. Specifically, we look at all subsets that are mergeable to a single part, perform {\em two} levels of recursive encoding and decoding, and measure the reconstruction error. The merge sequence with the lowest reconstruction error is added to the encoding hierarchy. The process repeats until no further merges are possible.
Particular cases of interest are {\em adjacency before symmetry}, and {\em symmetry before adjacency}, as illustrated in Figure \ref{fig:encode_order}. For each such case, we decode the final code back to the input box parameters (using, as for training, the known merging hierarchy) and measure the reconstruction error. This two-step lookahead is employed only for inferring hierarchies in test mode. During training, we minimize reconstruction loss over the hierarchy for the entire shape, as well as over all subtrees. Thus, the encoder/decoder units are tuned for both locally and globally good reconstructions, and at test time a relatively short lookahead suffices.

To decode a root code (e.g. one obtained from an encoding hierarchy inferred in the above fashion), we recursively invoke \NodeCat to decide whether which decoder should expand the node. The corresponding decoder (\AdjDec, \SymDec or \BoxDec) is used to recover the codes of child nodes until the full hierarchy has been expanded to leaves with corresponding box parameters. The test decoding setup is illustrated in Figure \ref{fig:ae_test}.

\begin{figure}[t] \centering
	\begin{overpic}[width=1.0\columnwidth,tics=100]{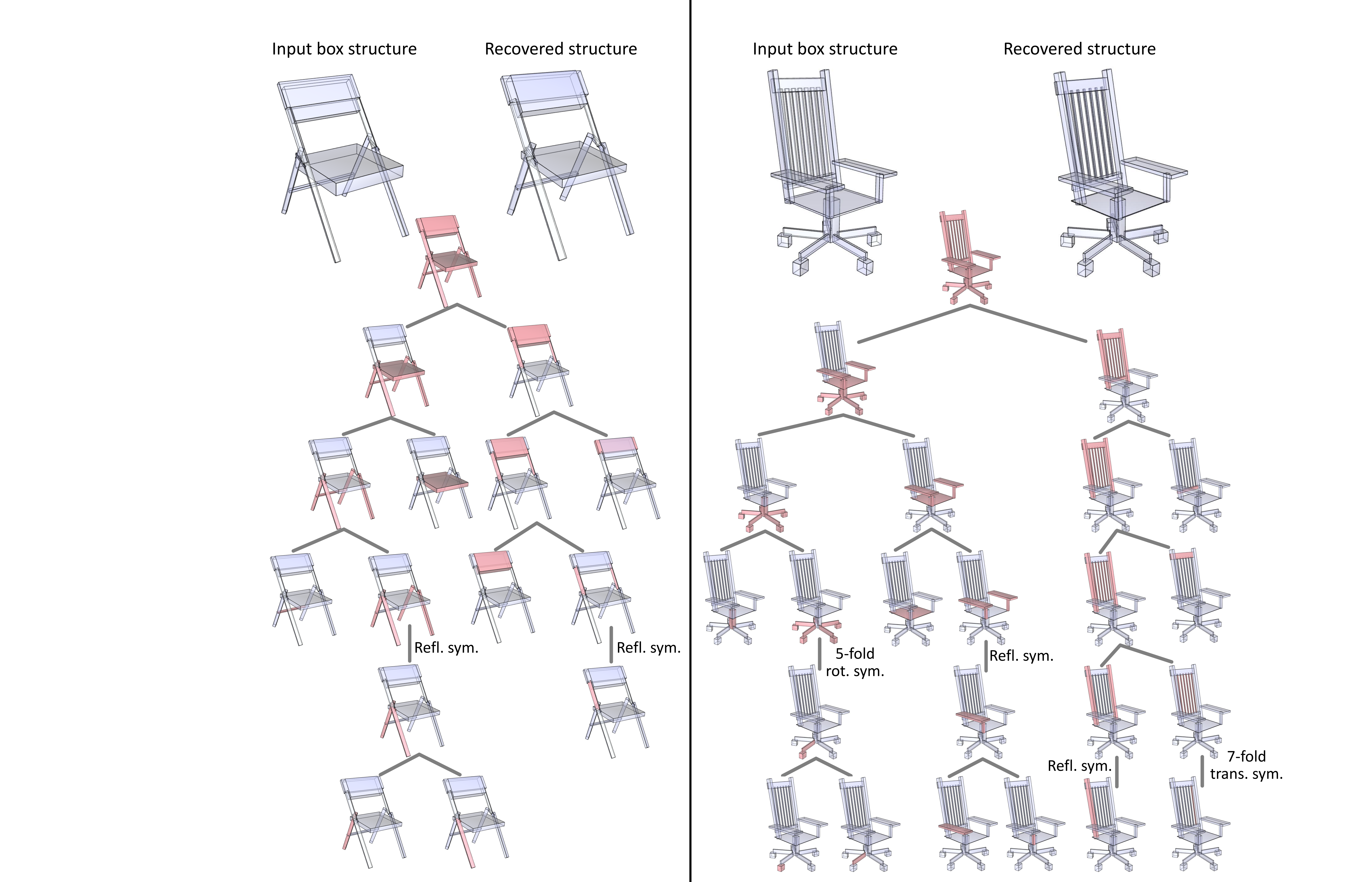}
	\end{overpic}
    \caption{Examples of reconstructing test shapes, without known hierarchies, by successively encoding them to root codes, and decoding them back. The encoding hierarchies inferred by our RvNN encoder are shown at the bottom.}
    \label{fig:test_recon}
\end{figure}

\begin{figure}[t] \centering
	\begin{overpic}[width=1.0\columnwidth,tics=100]{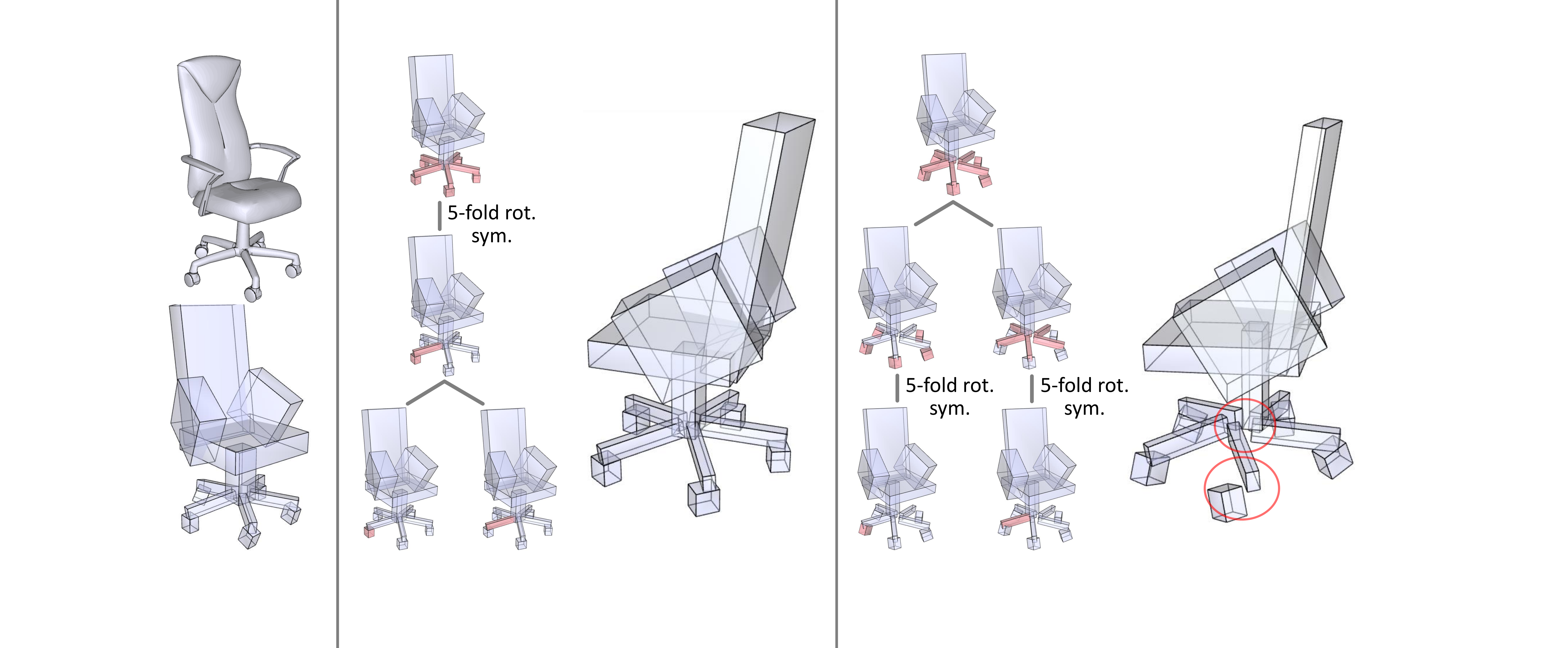}
    \put(5,-2.0){\small (a)}
    \put(34,-2.0){\small (b)}
    \put(78,-2.0){\small (c)}
	\end{overpic}
    \caption{Our RvNN encoder can find a perceptually reasonable symmetry hierarchy for a 3D shape structure, through minimizing reconstruction error. Given an input structure (a), the reconstruction error is much smaller if parts are grouped by adjacency before symmetry (b), instead of symmetry before adjacency (c).}
    \label{fig:symh_find}
\end{figure}

Several examples of test reconstructions are shown in Figure \ref{fig:test_recon}. The above procedures are used to encode a novel shape to a root code, and to reconstruct the shape given just this root code. In Figure~\ref{fig:symh_find}, we show how our RvNN is able to find a perceptually reasonable symmetry hierarchy for a 3D shape structure, by minimizing the reconstruction error. Given the structure of a swivel chair, the error is much smaller when a wheel and spike are merged before the 5-fold rotational symmetry is applied, than if two separate rotational symmetries (for wheels and spikes respectively) are applied first.

\section{Learning manifold of plausible structures}
\label{sec:gan}

Our recursive autoencoder computes a compact, fixed-dimensional code that represents the inferred hierarchical layout of shape parts, and can recover the layout given just this code associated with the root of the hierarchy. However, the autoencoder developed so far is not a generative model. It can reconstruct a layout from {\em any} root code, but an arbitrary, random code is unlikely to produce a plausible layout. A generative model must jointly capture the distribution of statistically plausible shape structures.

In this section, we describe our method for converting the auto-encoder-based model to a fully generative one. We fine-tune the autoencoder to learn a (relatively) low-dimensional manifold containing high-probability shape structures. Prior approaches for learning feasible manifolds of parametrized 3D shapes from landmark exemplars include kernel density estimation~\cite{Talton2009,Fish2014}, multidimensional scaling~\cite{Averkiou2014}, and piecewise primitive fitting~\cite{Shulz2016}. However, these methods essentially reduce to simple interpolation from the landmarks, and hence may assign high probabilities to parameter vectors that correspond to implausible shapes~\cite{goodfellow2014}.

Recently, {\em generative adversarial networks} (GANs)~\cite{goodfellow2014} have been introduced to overcome precisely this limitation. Instead of directly interpolating from training exemplars, a GAN trains a synthesis procedure to map arbitrary parameter vectors only to vectors which a classifier deems plausible. The classifier, which can be made arbitrarily sophisticated, is jointly trained to identify objects similar to the exemplars as plausible, and others as fake. This leads to a refined mapping of the latent space since implausible objects are eliminated by construction.
Given a completely random set of parameters, the trained GAN ``snaps'' it to the plausible manifold to generate a meaningful sample.

In addition to enabling the synthesis of novel but statistically plausible shape structures, the learned manifold also supports interpolation between shape codes. The application of this feature to shape morphing is shown in Section \ref{sec:results}.

\begin{figure}[t] \centering
	\begin{overpic}[width=1.0\columnwidth,tics=100]{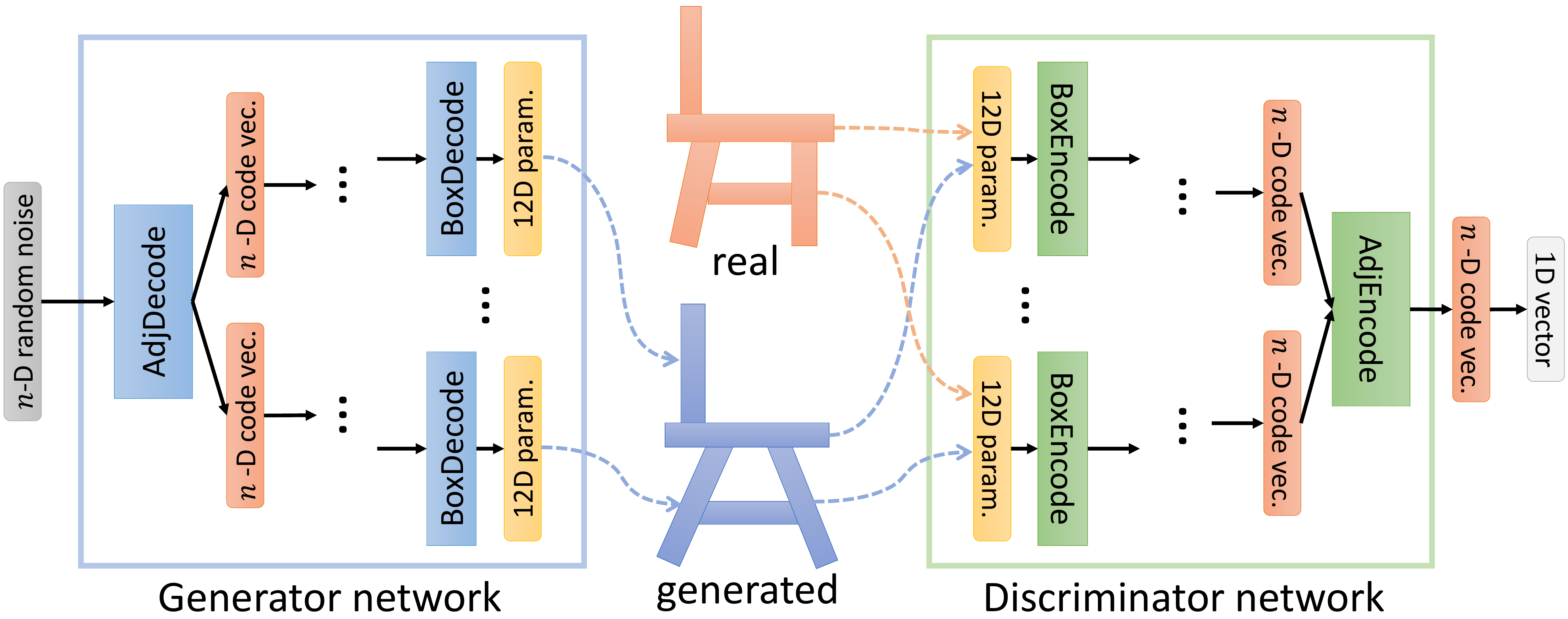}
	\end{overpic}
    \caption{Architecture of our generative adversarial network, showing reuse of autoencoder modules.}
    \label{fig:gan}
\end{figure}

\mypara{GAN architecture.} The architecture of our generative adversarial network comprises a generator ($G$) network, which transforms a random code to a hierarchical shape structure lying on the estimated manifold, and a discriminator ($D$) network, which checks whether a generated structure is similar to those of the training shapes or not. Our key observation is that we can {\em directly reuse and fine-tune the autoencoder modules} learned in the previous section, instead of introducing new components. The decoder component (comprising \AdjDec, \SymDec, \BoxDec and \NodeCat) is exactly what we need to estimate a structure from a given code: it constitutes the $G$ network. The encoder component (comprising \AdjEnc, \SymEnc and \BoxEnc) is exactly what we need to estimate a code for the generated structure. The final code can be compared to the codes of training structures using an additional fully connected layer and a binary softmax layer producing the probability of the structure being ``real''.
This constitutes the $D$ network. Hence, we initialize the GAN with the trained autoencoder modules and further fine-tune them to minimize the GAN loss. The architecture is illustrated in Figure \ref{fig:gan}, and the training procedure described below.

\mypara{Training.} The GAN is trained by stochastic gradient descent using different loss functions for the discriminator $D$ and the generator $G$. In each iteration, we sample two mini-batches: training box structures $x$ with their associated hierarchies, and random codes $z \in \R^n$. The $x$ samples, with known hierarchies, are passed only through the discriminator, yielding $D(x)$, whereas the $z$ samples are passed through both networks in sequence, yielding $D(G(z))$. The loss function for the discriminator is
\[
  J_D \ = \ -\frac{1}{2} E_x \left[\log D(x)\right] - \frac{1}{2} E_z \left[ \log(1 - D(G(z))) \right],
\]
while the loss function for the generator is
\[
  J_G \ = \ -\frac{1}{2} E_z \left[ \log D(G(z)) \right].
\]
By minimizing the first loss function w.r.t. the weights of the network $D$, we encourage the discriminator to output 1 for each training sample, and 0 for each random sample. By minimizing the second loss function w.r.t. the weights of the network $G$, we encourage the generator to fool $D$ into thinking a random sample is actually a real one observed during training. This is a standard adversarial training setup; see Goodfellow et al.~\shortcite{goodfellow2014} for more details.

\begin{figure}[t] \centering
	\begin{overpic}[width=1.0\columnwidth,tics=100]{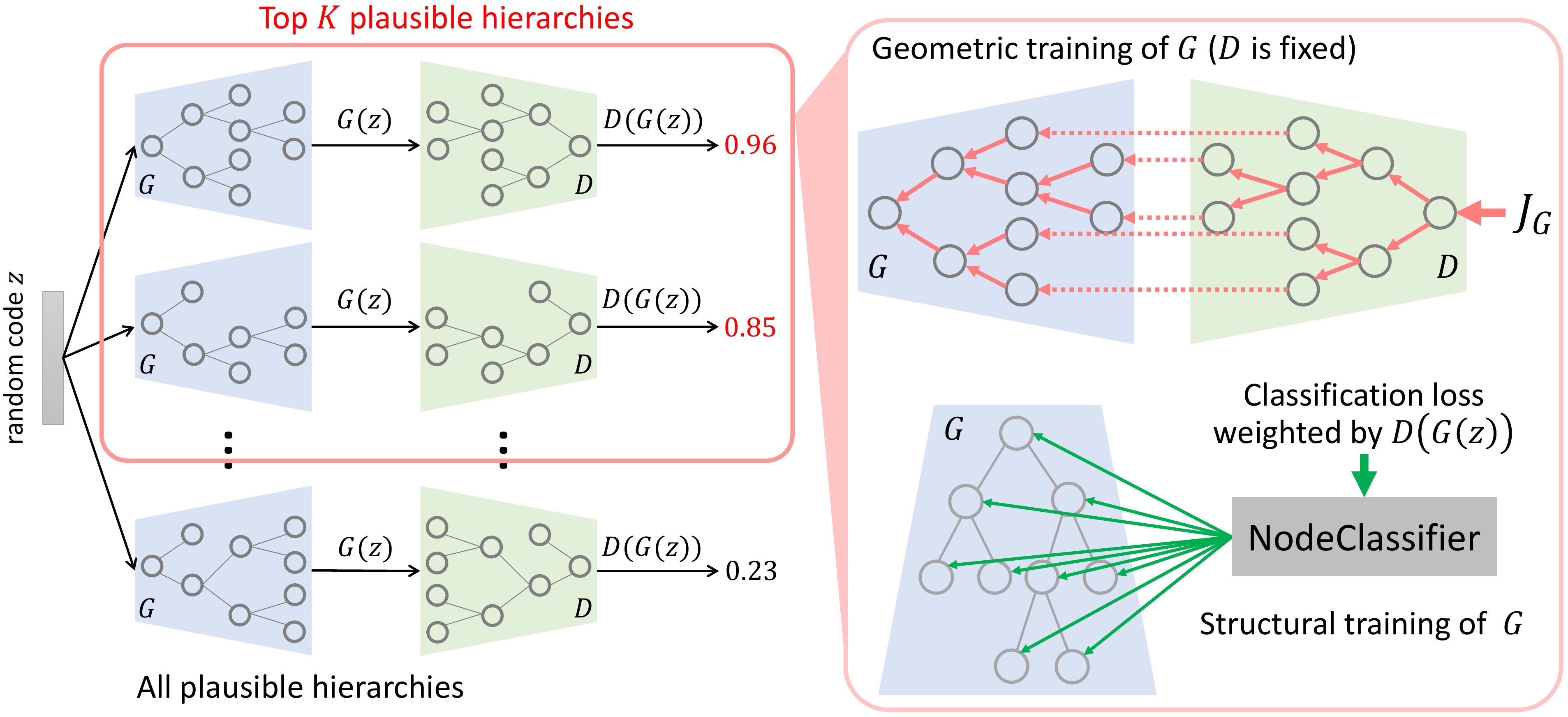}
	\end{overpic}
    \caption{The training of our GAN model. Left: Given a random code, we select the top $K$ ``plausible'' hierarchies from which $G$ can decode a box structure to best fool $D$.
    Right: For each selected hierarchy, the training of $G$ is split into geometric (top) and structural (bottom) tuning, based on different loss functions.}
    \label{fig:gan_train}
\end{figure}

\begin{figure}[t] \centering
	\begin{overpic}[width=1.0\columnwidth,tics=100]{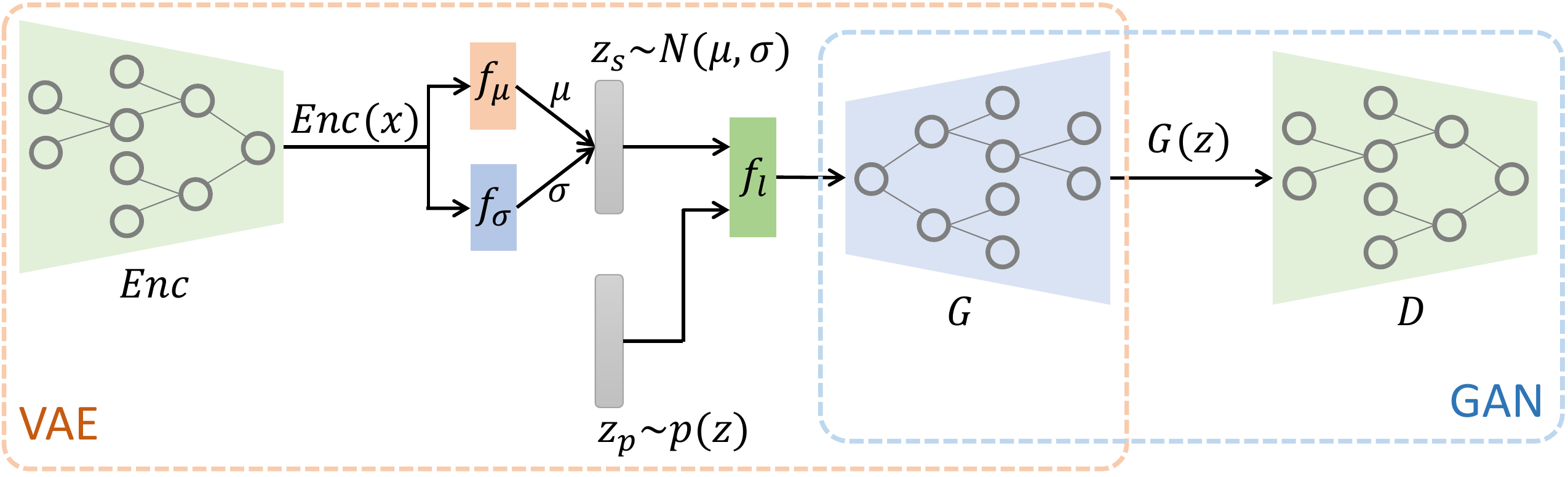}
	\end{overpic}
    \caption{Confining random codes by sampling from a learned Gaussian distributions based on learned root codes $Enc(x)$.
    Jointly learning the distribution and training the GAN leads to a VAE-GAN network.}
    \label{fig:vaegan}
\end{figure}

With this straightforward training, however, it is still hard to converge to a suitable balance between the $G$ and $D$ networks, despite the good initialization provided by our autoencoder. This is due to the following reasons.
\emph{First}, when mapping a random code $z$ to the manifold, the $G$ network (which is just the recursive decoder) may infer a grossly incorrect hierarchy.
The $D$ network finds it easy to reject these implausible hierarchies, and hence does not generate a useful training signal for $G$.
\emph{Second}, the implausible hierarchies generated from random codes may not provide reasonable pathways to back-propagate the loss from $D$ so that $G$ can be tuned properly.
\emph{Third}, since the decoding networks in $G$ are split into geometric (e.g. \AdjDec) and topological (\NodeCat) types,
they should be tuned separately with different losses deduced from $D$.
To these ends, we devise the following training strategies and priors, to better constrain the training process:
\begin{itemize}

  \item \emph{Structure prior for $G$.}
    In an initial stage, we need to prevent $G$ from mapping a random code $z$ to a severely implausible hierarchy. This is achieved by introducing a strong structure prior to $G$. We constrain the hierarchies inferred by $G$ to lie in a plausible set. This set includes all hierarchies used to train the autoencoder in Section \ref{sec:rvnn}. It also includes all hierarchies inferred by the autoencoder, in test mode, for the training shapes. For each $z$, we search the plausible hierarchies for the top $K=10$ ones that best fool the discriminator, minimizing $J_G$ (Figure~\ref{fig:gan_train}, left). These hierarchies are then used to back-propagate the loss $J_G$.
    \smallskip

  \item \emph{Separate geometric and structural training.}
    Given a selected hierarchy, we first tune the geometric decoders of $G$ via back-propagating the corresponding loss $J_G$ through the hierarchy. This tuning is expected to further fool the discriminator, leading to a higher estimate $D(G(z))$ that $G(z)$ is real (Figure~\ref{fig:gan_train}, top-right). For each selected hierarchy, with its the newly updated $D(G(z))$, we then tune the structural component, \NodeCat, of $G$. This is done by minimizing the classification loss of \NodeCat at each node in the given hierarchy, using the node type as ground-truth (Figure~\ref{fig:gan_train}, bottom-right). To favor those hierarchies that better fool $D$, we weight the loss by $D(G(z))$.
    \smallskip

  \item \emph{Constrained random code sampling.}
    Given the priors and constraints above, it is still difficult to train $G$ to reconstruct a plausible hierarchy from arbitrarily random codes. Therefore, instead of {\em directly} feeding it random codes from a normal or uniform distribution, we feed it codes drawn from Gaussians around the training samples $x$, whose {\em mixture} approximates the standard normal distribution. Further, we train a secondary network $f_l$ to project these ``latent'' codes to a space of potential root codes which are easier for $G$ to process.

    Specifically, $G$ takes samples from a multivariate Gaussian distribution: $z_s(x) \sim N(\mu,\sigma)$ with $\mu=f_{\mu}(Enc(x))$ and $\sigma=f_{\sigma}(Enc(x))$. Here, $Enc$ is the recursive encoder originally trained with the autoencoder (before adversarial tuning), running in test mode. $f_{\mu}$ and $f_{\sigma}$ can be approximated by two neural networks.

    We train to minimize the reconstruction loss on $x$, in addition to the generator loss in the GAN. In fact, the networks $Enc$ and $G$ constitute a variational autoencoder (VAE) if we also tune $Enc$ when learning the parameters of the Gaussian distribution. This leads to an architecture similar to the VAE-GAN proposed by Larsen et al.~\shortcite{larsen2015}; see Figure~\ref{fig:vaegan}.

    Consequently, we also impose the loss function for VAE that pushes this variational distribution $p(z_s(x))$, over all training samples $x$, towards the prior of the standard normal distribution $p(z)$. In summary, we minimize the following loss function:
    \begin{equation*}\label{eq:vaeganloss}
    L = L_{\text{GAN}}(z_p) + \alpha_1 L_{\text{recon}} + \alpha_2 L_{\text{KL}}
    \end{equation*}
    The GAN loss is $L_{\text{GAN}} = \log D(x) + \log(1 - D(G(f_l(z_p))))$, with $z_p \sim p(z)$. This loss is minimized/maximized by $G$/$D$, respectively.
    The reconstruction loss is defined as $L_{\text{recon}} = \|G(f_l(z_s(x))) - x\|_2$.
    The Kullback-Leibler divergence loss, $L_{\text{KL}} = D_{\text{KL}} \left( p(z_s(x)) ~\|~ p(z) \right)$, forces the mixture of local Gaussians to approximate the standard normal distribution. We set $\alpha_1=10^{-2}$ and $\alpha_2=10$ in our experiments.
\end{itemize}

The results of the GAN training process are fine-tuned RvNN decoder modules and the $f_l$ network. The $f_l$ network projects a random \mbox{$n$-D} vector drawn from the standard normal distribution to the space of potential root codes, and the tuned decoders map the projected vector to a structure lying on the plausible manifold. Together with a module to generate fine-grained part geometry, described in the next section, this constitutes our recursive, generative model of 3D shapes.

\section{Part geometry synthesis}
\label{sec:partgeom}

In the previous sections, we described our generative model of part layouts in shapes. The final component of our framework is a generative model for fine-grained part geometry, conditioned on the part bounding box and layout. Our solution has two components. First, we develop a fixed-dimensional part feature vector that captures both the part's gross dimensions and its context within the layout. Second, we learn a low-dimensional manifold of plausible part geometries while simultaneously also learning a mapping from part feature vectors to the manifold. This mapping is used to obtain the synthesized geometry for a given part in a generated layout. Below, we describe these steps in detail.

\begin{figure}[b] \centering
	\begin{overpic}[width=1.0\columnwidth,tics=100]{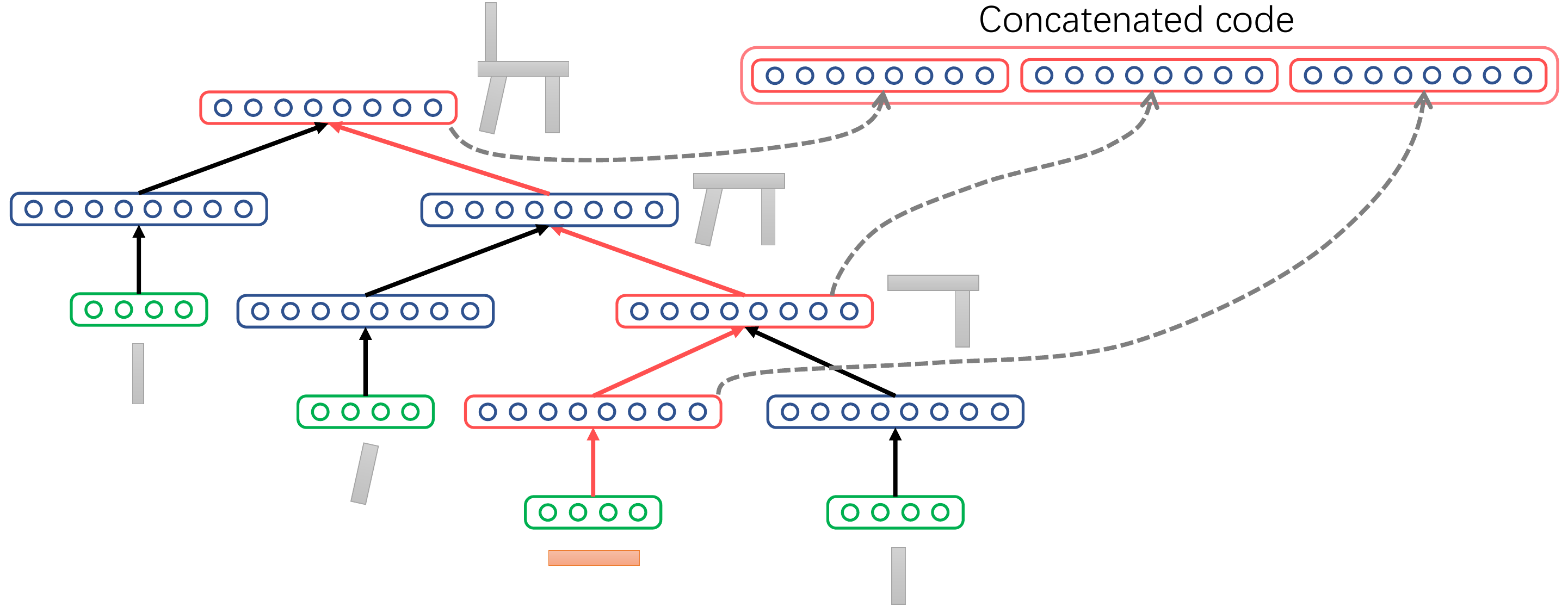}
	\end{overpic}
    \caption{Construction of structure-aware recursive feature (SARF) for a part in a hierarchy. We concatenate the RvNN codes of the part, its immediate parent, and the root into a fixed-dimensional vector.}
    \label{fig:sarf}
\end{figure}

\mypara{Structure-aware recursive feature (SARF)} The recursive generator network produces a hierarchy of shape parts, with each internal node in the hierarchy represented by an \mbox{$n$-D} code. We exploit this structure to define a feature vector for a single part. A natural contextual feature would be to concatenate the RvNN codes of all nodes on the path from the part's leaf node to the root. However, since paths lengths are variable, this would not yield a fixed-dimensional vector. Instead, we approximate the context by concatenating just the code of the leaf node, that of its immediate parent, and that of the root into a \mbox{$3n$-D} feature vector (Figure \ref{fig:sarf}). The first code captures the dimensions of the part's bounding box, and the latter two codes capture local and global contexts, respectively.

\begin{figure}[t] \centering
	\begin{overpic}[width=1.0\columnwidth,tics=100]{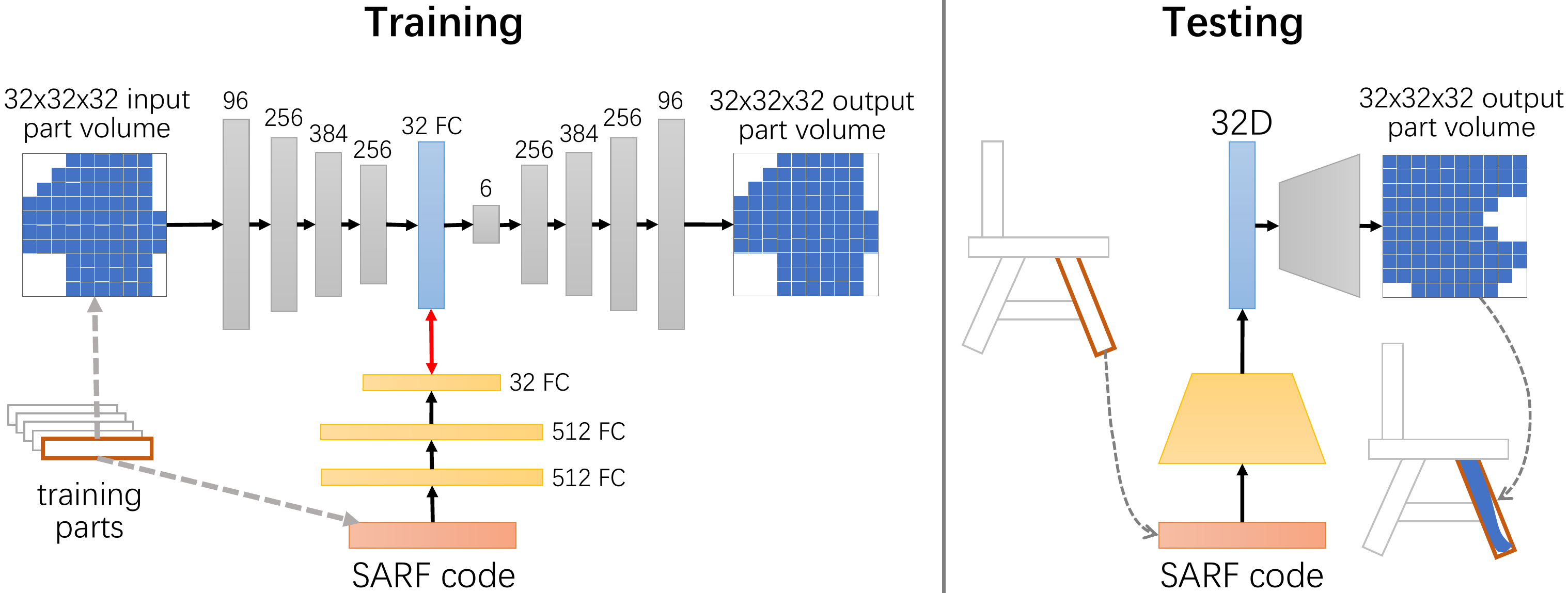}
	\end{overpic}
    \caption{Training and testing setup for part geometry synthesis.}
    \label{fig:geom_train_test}
\end{figure}

\begin{figure}[t] \centering
	\begin{overpic}[width=1.0\columnwidth,tics=100]{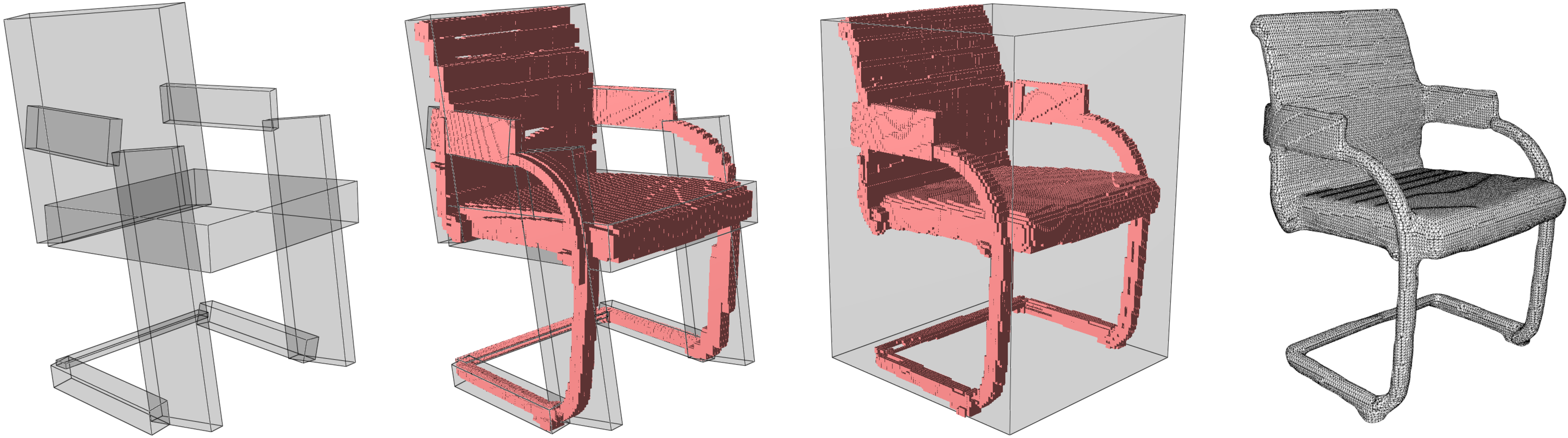}
    \put(12,-3){\small (a)}
    \put(36,-3){\small (b)}
    \put(63,-3){\small (c)}
    \put(88,-3){\small (d)}
	\end{overpic}
    \caption{Geometry synthesis from part structure. Given a generated part structure (a), we synthesize the geometry inside each part box in volumetric representation (b).
    The per-box volumes are then embedded into a global volume (c) from which we reconstruct the final meshed model (d).}
    \label{fig:geom_synth}
\end{figure}

\mypara{SARF to part geometry.} In the second stage, we would like to map a SARF feature vector to the synthesized geometry for the part, represented in our prototype as a \mbox{$32 \times 32 \times 32$} voxel grid. Such a mapping function is difficult to train directly, since the output is very high (8000) dimensional yet the set of {\em plausible} parts spans only a low-dimensional manifold within the space of all outputs. Instead, we adapt a strategy inspired by Girdhar et al.~\shortcite{girdhar2016}. We set up a deep, convolutional autoencoder, consisting of an encoder \GeoEnc to map the voxel grid to a compact, 32D part code, and a decoder \GeoDec to map it back to a reconstructed grid. The learned codes efficiently map out the low-dimensional manifold of plausible part geometries. We use the architecture of Girdhar et al., and measure the reconstruction error as a sigmoid cross-entropy loss. Simultaneously, we use a second deep network \GeoMap to map an input SARF code to the 32D part code, with both networks accessing the same code neurons. The mapping network employs a Euclidean loss function. We train both networks jointly, using both losses, with stochastic gradient descent and backpropagation. At test time, we chain together the mapping network \GeoMap and the decoder \GeoDec to obtain a function mapping SARF codes to synthesized part geometry. The training and test setups are illustrated in Figure \ref{fig:geom_train_test}. The synthesis of the overall shape geometry is done by predicting part-wise 3D volumes, which are then embedded into a global volume, from which we reconstruct the final meshed model. See Figure~\ref{fig:geom_synth} for an example.

\section{Results and evaluation}
\label{sec:results}

We evaluate our generative recursive model of shape structures through several experiments. First, we
focus on validating that our autoencoder-based RvNN learns the ``correct'' symmetry hierarchies, where
correctness could be qualified in different ways, and the resulting codes are useful in applications such as
classification and partial matching. Then we test the generative capability of our VAE-GAN network
built on top of the RvNN.


\mypara{Dataset:} We collected a dataset containing $1000$ 3D models from five shape categories: chairs ($500$), bikes ($200$), aeroplanes ($100$), excavators ($100$), and candelabra ($100$). These models are collected from the ShapeNet and the Princeton ModelNet. Each model is pre-segmented according to their mesh components or based on the symmetry-aware segmentation utilized in~\cite{wang2011}. The average number of segments per shape is $12$ for chairs, $10$ for bikes, $7$ for aeroplanes, $6$ for excavators, and $8$ for candelabra. Symmetric parts are counted as distinct. We do not utilize any segment labels.

Our RvNN autoencoder is trained with all shapes in the dataset. The generative VAE-GAN is trained per category, since its training involves structure learning which works best within the same shape category. Part geometry synthesis is trained on all parts from all categories.

\begin{figure}[!t] \centering
	\begin{overpic}[width=1.0\columnwidth,tics=100]{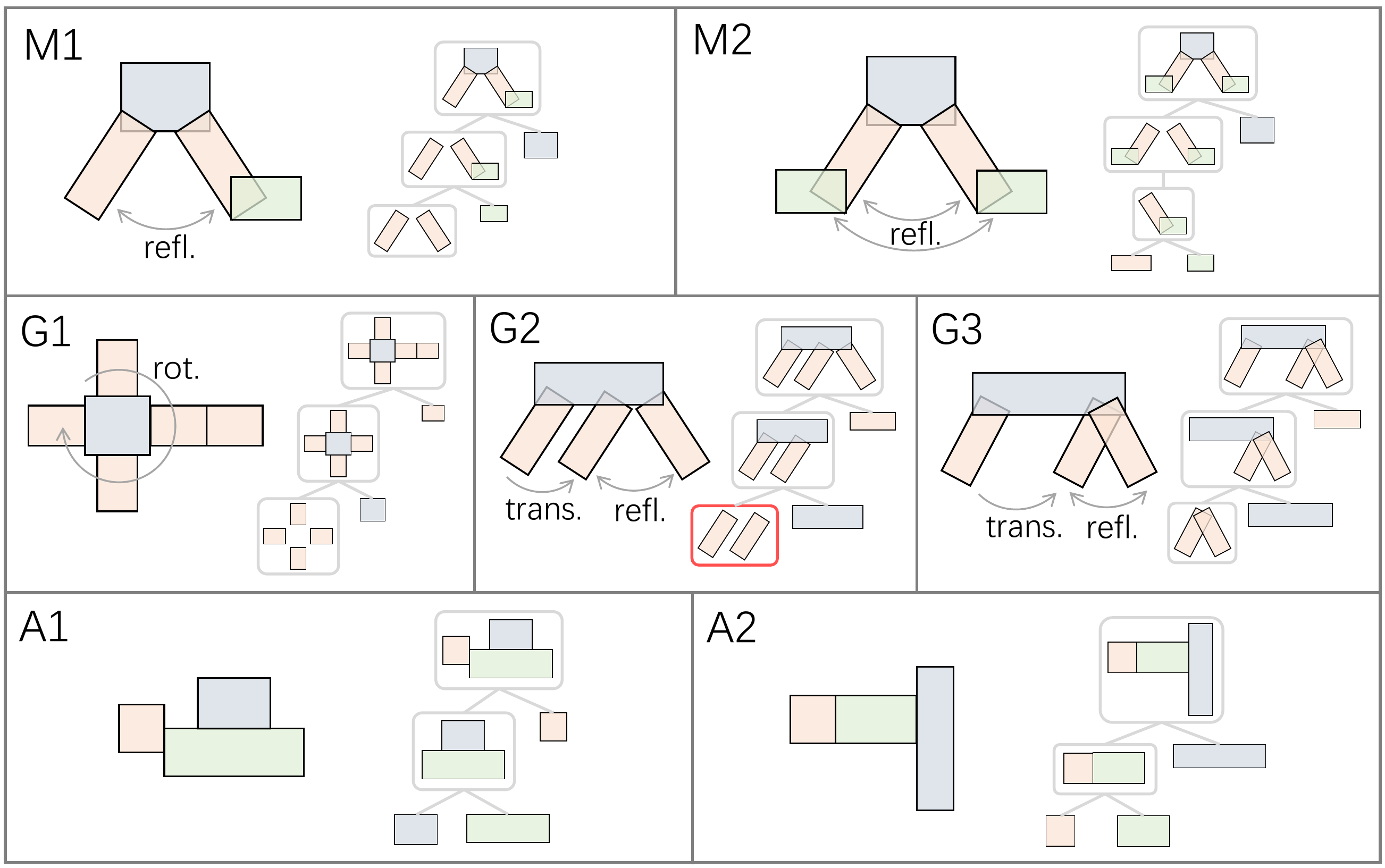}
	\end{overpic}
    \caption{Our RvNN encoder correctly parses six out of seven 2D box arrangements designed to test handcrafted, perceptually-based grouping rules from~\protect\cite{wang2011}. The G2 rule is violated in our example, with 2-fold translational symmetry (highlighted in the red box) taking precedence over the reflectional one.}
    \label{fig:SYMH-rules}
\end{figure}

\mypara{Learning recursive grouping rules.}
In the original work on symmetry hierarchies by Wang et al.~\shortcite{wang2011}, a total of seven precedence rules
(labeled M1, M2, G1, G2, G3, A1, and A2; see the Appendix for a reproduction of these rules)
were handcrafted to determine orders between and among assembly and
symmetry grouping operations. For example, rule A1 stipulates that symmetry-preserving assembly should
take precedence over symmetry-breaking assembly and rule M2 states that assembly should be before
grouping (by symmetry) if and only if the assembled elements belong to symmetry groups which possess equivalent
grouping symmetries. These rules were inspired by Gestalt laws of perceptual grouping~\cite{Koehler1929} and
Occam's Razor which seeks the simplest explanation. One may say that they are perceptual and represent a
certain level of human cognition.

The intriguing question is whether our RvNN, which is unsupervised, could ``replicate'' such cognitive capability.
To test the rules, we designed seven box arrangements in 2D, one per rule; these patterns are
quite similar to those illustrated in Wang et al.~\shortcite{wang2011}. For rule A2, which involves
a connectivity strength measure, we simply used geometric proximity.
%
%
In Figure~\ref{fig:SYMH-rules}, we show the seven box arrangements and the grouping learned by our RvNN.
%
As can be observed,  our encoder correctly parses all expected patterns except in the case of G2, where 2-fold translational symmetry takes precedence over the reflectional one in our example.

\begin{figure}[!t] \centering
	\begin{overpic}[width=1.0\columnwidth,tics=100]{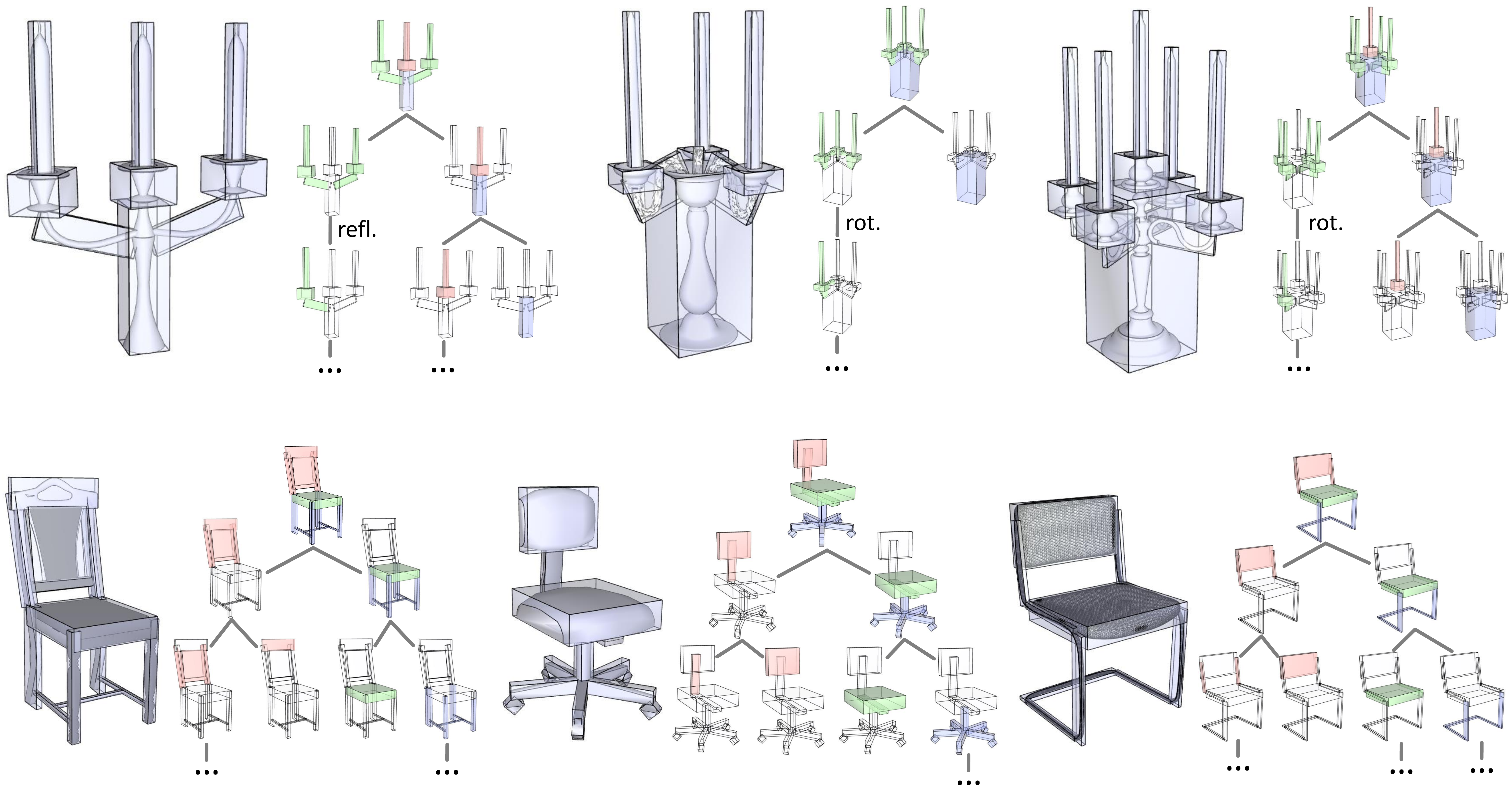}
	\end{overpic}
    \caption{Inferred hierarchies are consistent across sets of shapes, shown for two shape classes (candelabra and chairs).}
    \label{fig:consistency}
\end{figure}

\mypara{Consistency of inferred hierarchies.} Our RvNN framework infers hierarchies consistently across different shapes. To demonstrate this, we augment two categories of our segmented dataset -- {\em chair} and {\em candelabra} -- with semantic labels (e.g., for chairs: ``seat'', ``back'', ``leg'', and ``arm''). Note that these labels occur at relatively higher levels of the hierarchies, since legs, backs, etc., may be subdivided into smaller parts. If the hierarchies are consistent across shapes, these high-level labels should follow a consistent merging order. For example, the seat and legs should be merged before the seat and back are merged. Let $\ell_p$ denote the label of part $p$. Given another label $\ell$, let $h(p, \ell)$ denote the shortest distance from $p$ to an ancestor that it shares with a part with label $\ell$. Note that $h(p, \ell_p) = 0$ by definition. Let $S_\ell$ be the set of parts with label $\ell$. For labels $\ell_1, \ell_2$, we measure the probability $P_\ell(\ell_1 \prec \ell_2)$ that $\ell_1$ is more regularly grouped with $\ell$ than $\ell_2$ as $\sum_{p \in S_\ell} \mathbb{I}(h(p, \ell_1) < h(p, \ell_2)) / |S_\ell|$, where $\mathbb{I}$ is the indicator function and the sum is additionally restricted over shapes in which all three labels appear. The overall consistency is estimated as one minus the average entropy over all label triplets:
\[
C \ = \ 1 \ + \ \binom{|L|}{3}^{-1} \sum_{\ell, \ell_1, \ell_2 \in L,\ \ell \neq \ell_1 \neq \ell_2} P_\ell(\ell_1 \prec \ell_2) \log_2 P_\ell(\ell_1 \prec \ell_2)
\]
The average consistency over the two categories of training shapes was measured as $0.81$, and over the two categories of test shapes as $0.72$. The high values show that our RvNN infers hierarchies consistently across different shapes. Figure \ref{fig:consistency} shows several pairs of shapes with consistent inferred hierarchies.

\mypara{Classification of shape structures.}
Our autoencoder generates compact encodings for shapes segmented into arbitrary numbers of parts, via a recursively inferred hierarchy. To test whether these codes effectively characterize shapes and shape similarities, we conducted a {\em fine-grained\/} shape classification experiment for each of four classes: airplane, chair, bike, and candle. The sub-classes
were: airplane -- $5$ classes including jet, straight-wing, fighter, delta-wing, swept-wing; chair -- $5$ classes including armchair, folding, swivel, four-leg, sofa; bike -- $4$ classes including
motorcycle, casual bicycle, tricycle, mountain bike; and candelabra -- $3$ classes including with arms, w/o arm, with two-level arms. To represent each shape, we used the average of all codes in the shape's hierarchy, which, as in Socher et al.~\shortcite{socher2011}, we found to work better than just the root.


Following the standard protocol for each category of shapes, we hold out one shape in turn, and sort the remaining shapes by increasing the $L_2$ distance between average codes, terminating the results by a variable upper limit on the distance. The number of results from the class of the query shape are considered as true positives.

\begin{figure}[!t] \centering
	\begin{overpic}[width=1.0\columnwidth,tics=100]{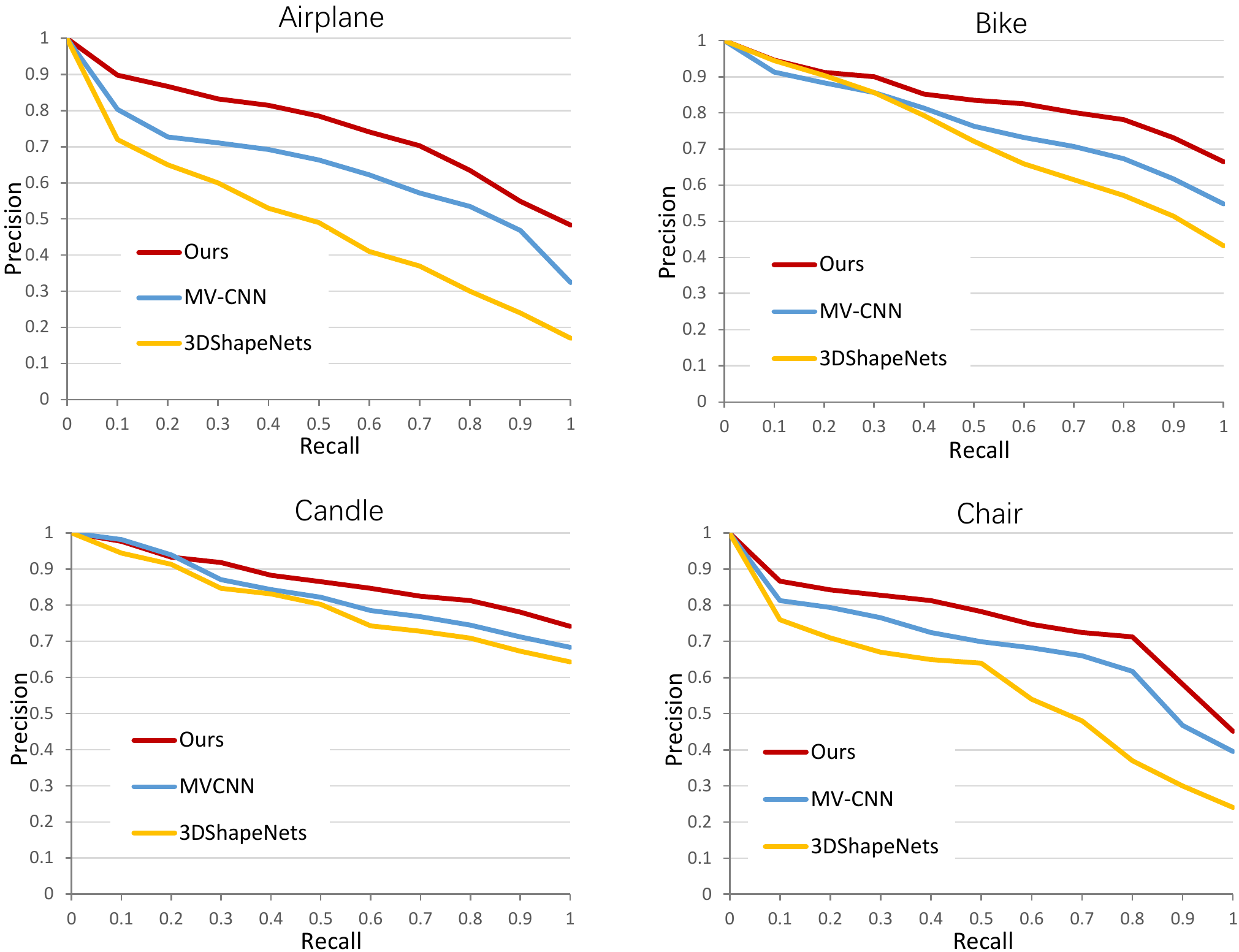}
	\end{overpic}
    \caption{Precision-recall plots for classification tasks.}
    \label{fig:retrieval}
\end{figure}

We show precision-recall plots for four classes of interest in Figure \ref{fig:retrieval}. The average accuracy of (subclass) classification over all four classes is $96.1\%$.

As baselines, we show the performance of two state-of-the-art descriptors on this task~\cite{wu2015,su2015}. This is not an entirely equal-grounded comparison: our method leverages a prior segmentation of each shape into (unlabeled) parts, whereas the baseline methods do not. However, our method does not consider any fine-grained part geometry, only oriented bounding box parameters. The considerable improvement of our method over the baselines demonstrates that gross structure can be significantly more important for shape recognition than fine-grained geometry, and accurate and consistent identification of part layouts can be the foundation of powerful retrieval and classification methods.

\begin{figure}[b] \centering
	\begin{overpic}[width=1.0\columnwidth,tics=100]{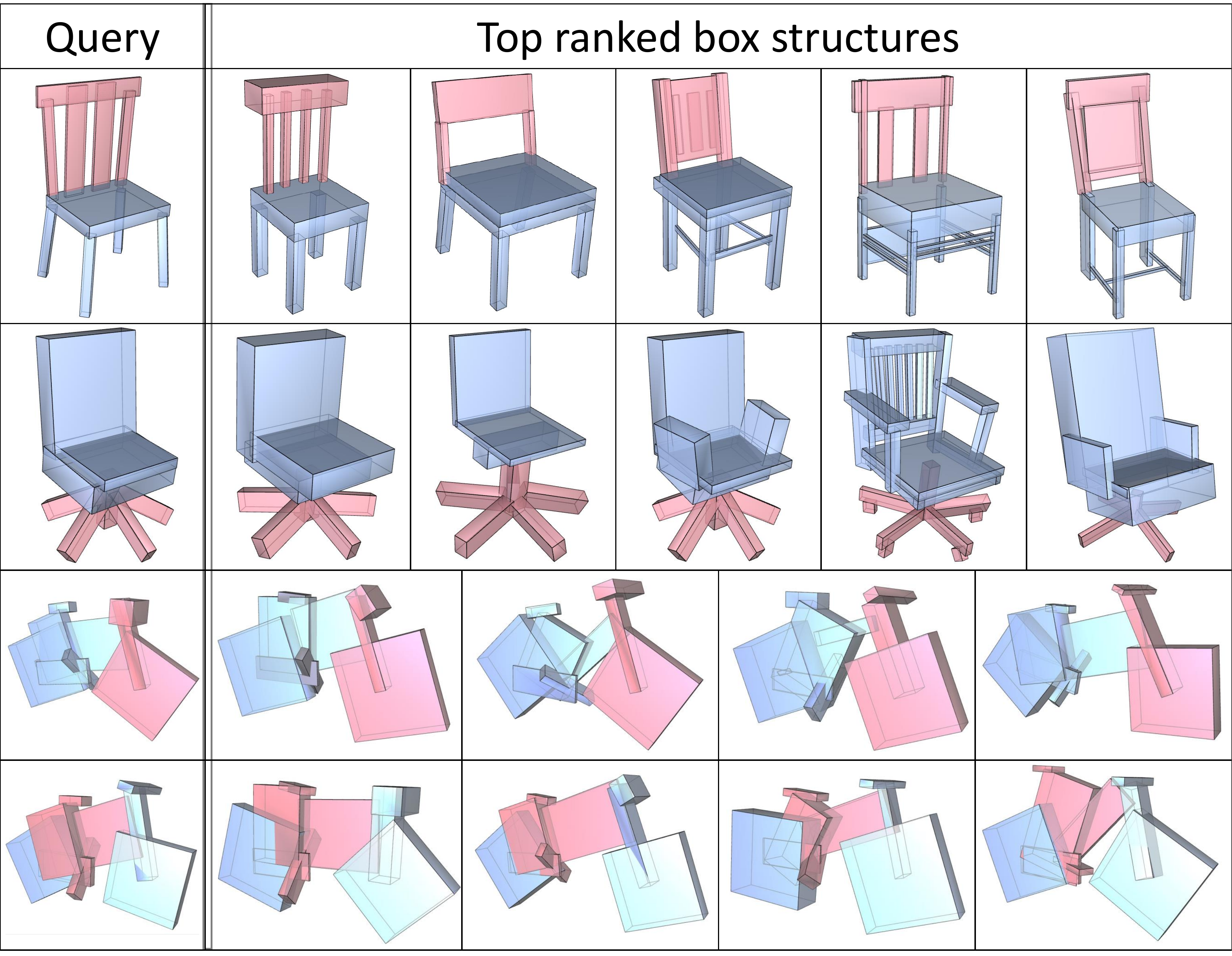}
	\end{overpic}
    \caption{Partial structure retrieval results for two shape classes (chair and bicycle). The query and matching parts are highlighted in red.
    }
    \label{fig:partial}
\end{figure}

\mypara{Partial structure matching.}
While the previous experiment tested full shape retrieval, it is also interesting to explore whether subtree codes are sufficiently descriptive for part-in-whole matching.
As before, we use the average of codes in a subtree as the feature for the subtree. Figure~\ref{fig:partial} contains some partial retrieval results, showing that our method correctly retrieves subparts matching the query.


\begin{figure}[t] \centering
	\begin{overpic}[width=1.0\columnwidth,tics=100]{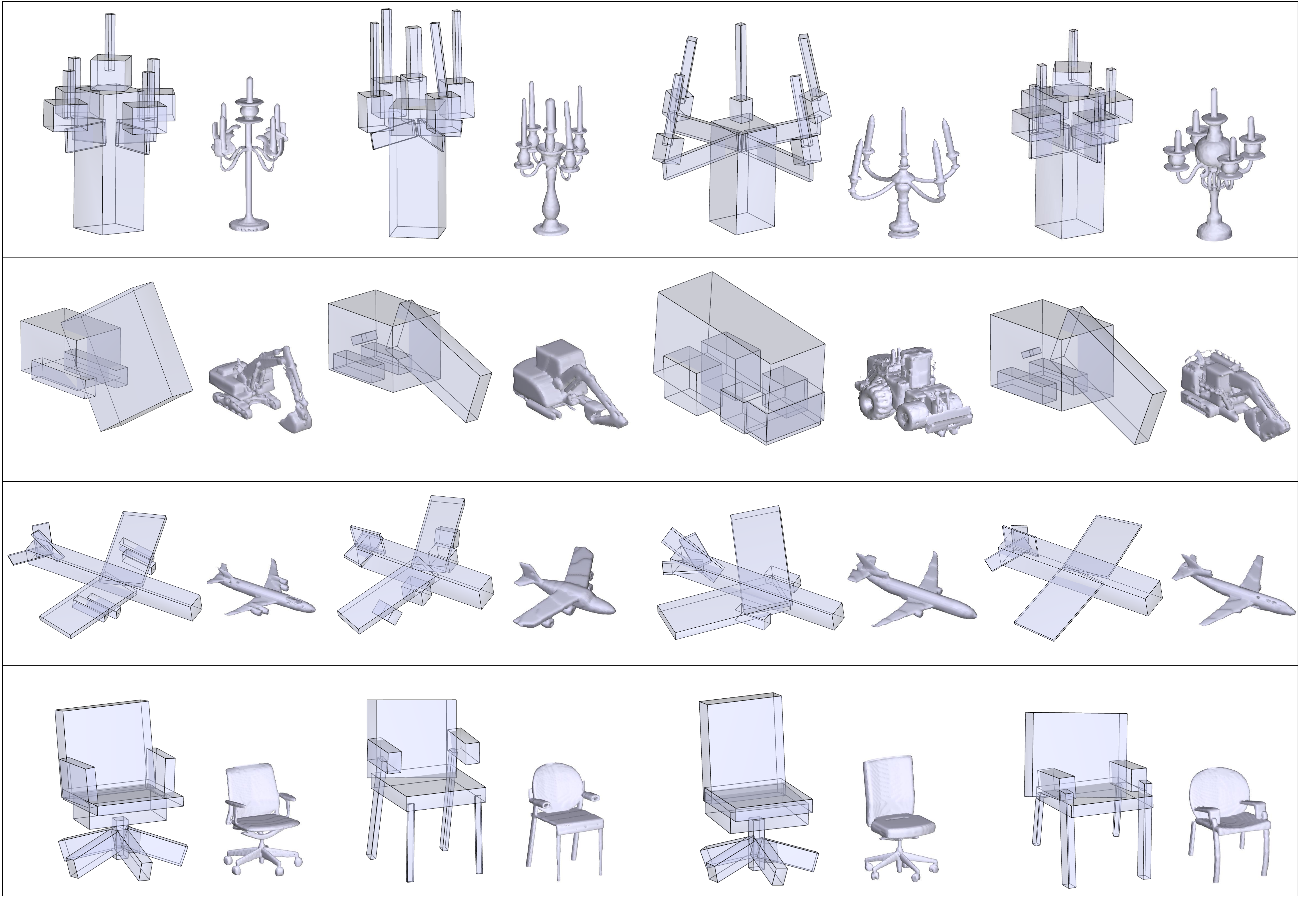}
	\end{overpic}
    \caption{Examples of shapes synthesized from different classes.}
    \label{fig:synthesis}
\end{figure}

\mypara{Shape synthesis and interpolation.} Our framework is generative, and can be used to synthesize shapes from the learned manifold in a two-step process. First, the VAE-GAN network is sampled using a random seed for a hierarchical bounding box layout. Second, the leaf nodes of the hierarchy are mapped to fine-grained voxelized geometry, which is subsequently meshed. Several examples of synthesized shapes are shown in Figure \ref{fig:synthesis}.

\begin{figure*}[h] \centering
	\begin{overpic}[width=0.9\linewidth,tics=100]{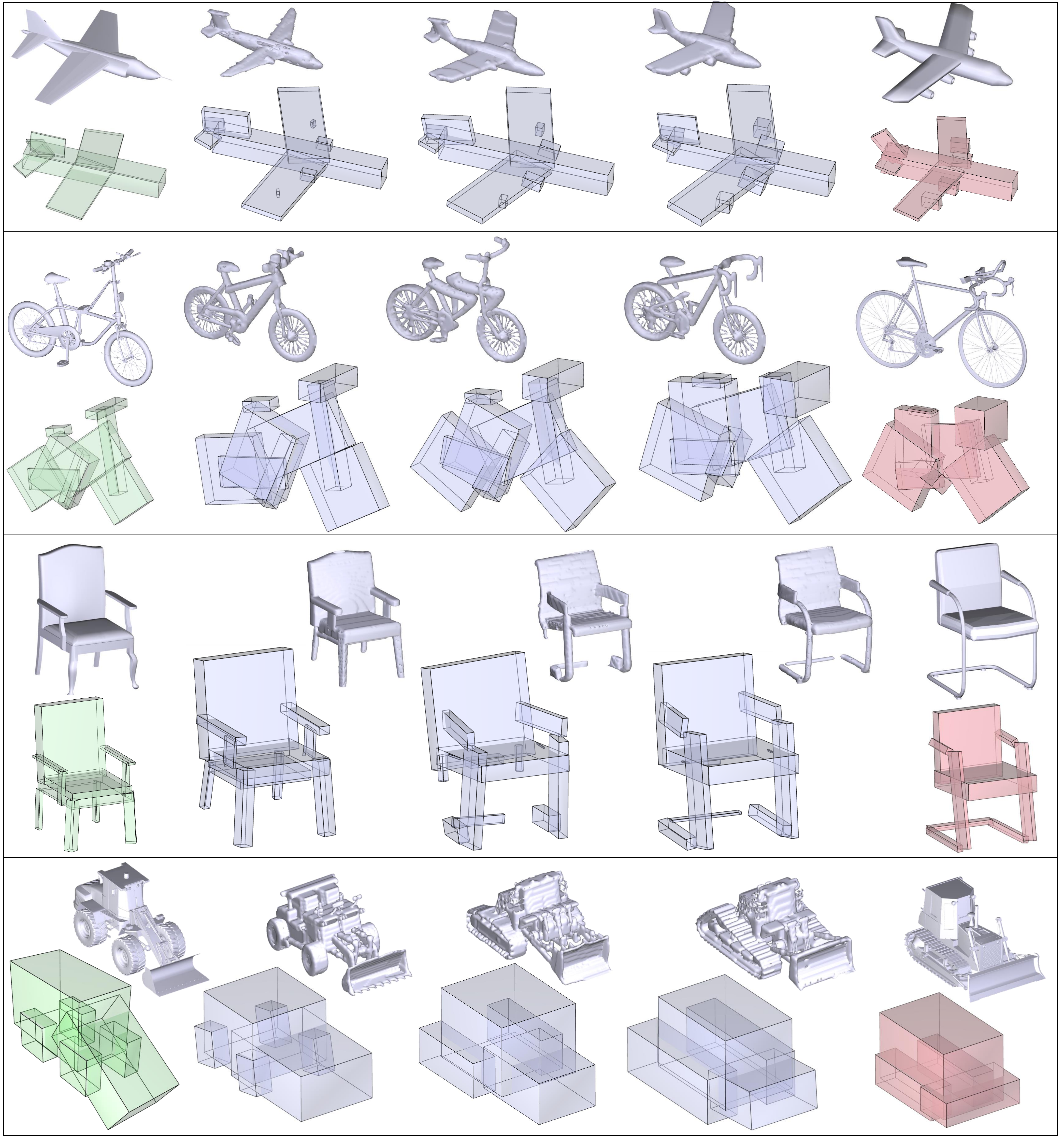}
	\end{overpic}
    \caption{Linear interpolation between root codes, and subsequent synthesis, can result in plausible morphs between shapes with significantly different topologies.}
    \label{fig:interpolation}
\end{figure*}

Our model can also be used to interpolate between two topologically and geometrically different shapes. For this task, we compute the root codes of two shapes via inferred hierarchies. Then, we linearly interpolate between the codes, reconstructing the shape at each intermediate position using the synthesis procedure above. Although intermediate codes may not themselves correspond to root codes of plausible shapes, the synthesis procedure projects them onto the valid manifold by virtue of the VAE-GAN training. We demonstrate example interpolations in Figure \ref{fig:interpolation}. Note that our model successfully handles topological changes both in the part layout and within parts, while maintaining symmetry constraints. Unlike Jain et al.~\shortcite{Jain2012}, we do not require prior knowledge of part hierarchies. Unlike both Jain et al. and Alhashim et al.~\shortcite{alhashim2014}, we do not require part correspondences either, and we can handle smooth topological changes in individual parts.

\mypara{Implementation and Timing.} Our RvNN and VAE-GAN are implemented in MATLAB.
The geometry synthesis model is implemented using the MatConvNet neural network library.
Pre-training the autoencoder (Section \ref{sec:rvnn}) took $14$ hours.
Adversarial fine-tuning (Section \ref{sec:gan}) took about $20$ hours for each shape class.
Training the part geometry synthesis network (Section \ref{sec:partgeom}) took $25$ hours.
Mapping a random code vector to the manifold of plausible structures to synthesize a hierarchy takes $0.5$ seconds,
and augmenting it with synthesized fine-grained part geometry takes an additional $0.2$ seconds per part.

\section{Discussion, limitation, and future work}
\label{sec:future}

With the work presented, we have only made a first step towards developing a structure-aware,
generative neural network for 3D shapes. What separates our method apart from previous attempts
at using neural nets for 3D shape synthesis is its ability to learn, without supervision, and
synthesize {\em shape structures\/}. It is satisfying to see that the generated 3D shapes
possess cleaner part structures, such as symmetries, and more regularized part geometries, when
compared to voxel fields generated by previous works~\cite{wu2016,girdhar2016}. What is unsatisfying
however is that we decoupled the syntheses of structure and fine geometry. This hints at an
obvious next step to integrate the two syntheses.

The codes learned by our RvNN do combine structural and geometric
information into a single vector. Through experiments, we have demonstrated that the
hierachical grouping learned by the RvNN appears to conform to perceptual principles as reflected
by the precedence rules handcrafted by Wang et al.~\shortcite{wang2011}.
The codes also enable applications such as fine-grained classification and partial shape retrieval,
producing reasonable results. However, the internal mechanisms of the code and precisely how
it is mixing the structural and geometric information is unclear. The fact that it appears to be
able to encode hiearachies of arbitrary depth with a fixed-length vector is even somewhat mysterious.
An interesting future work would be to ``visualize'' the code to gain an insight on all of these questions.
Only with that insight would we be able to steer the code towards a better separation between the parts
reflecting the structure and the parts reflecting low-level geometry.


Our current network still has a long way to go in fully mapping the {\em generative structure manifold\/}. We cannot extrapolate arbitrarily -- we are limited to a VAE-GAN setup which samples codes similar to, or in between, the exemplars. Hence, our synthesis and interpolation are confined to a local patch of that elusive ``manifold''. In fact, it is not completely clear whether the generative structure space for a 3D shape collection with sufficiently rich structural variations is a low-dimensional manifold. Along similar lines, we have not discovered flexible mechanisms to generate valid codes, e.g., by applying algebraic or crossover operations, from available codes. All of these questions and directions await future investigations.
It would be interesting to thoroughly investigate the effect of code length on structure encoding.
Finally, it is worth exploring recent developments in GANs, e.g. Wasserstein GAN~\cite{arjovsky2017}, in our problem setting. It would also be interesting to compare with plain VAE and other generative adaptations.



\section*{Acknowledgements}
We thank the anonymous reviewers for their valuable comments and suggestions.
We are grateful to Yifei Shi, Min Liu and Yizhi Wang for their generous help in data preparation and result production.
Jun Li is a visiting PhD student at the University of Bonn, supported by the China Scholarship Council.
This work was supported in part by
NSFC (61572507, 61532003, 61622212), an NSERC grant (611370),
NSF Grants IIS-1528025 and DMS-1546206,
a Google Focused Research Award,
and awards from the Adobe, Qualcomm and Vicarious corporations.

\bibliographystyle{acmsiggraph}
\bibliography{shape_dna}


\section*{Appendix: \\ Precedence rules for symmetry hierarchy}

We reproduce the precedence rules stipulated in Wang et al.~\shortcite{wang2011}
for sorting symmetry grouping and assembly operations:


\noindent
{\bf M1 (Grouping before assembly):}
Grouping by symmetry takes precedence over assembly operations, with an exception given by the next rule ({\bf M2}).

\noindent
{\bf M2 (Assembly before grouping):}
Assemble before grouping if and only if the assembled nodes belong to symmetry cliques which possess
equivalent grouping symmetries.



\noindent
{\bf G1 (Clique order):}
If there are still symmetry cliques of order greater than two in the contraction graph, then
higher-order cliques are grouped before lower-order ones.

\noindent
{\bf G2 (Reflectional symmetry):}
If there are only order-2 cliques in the graph, then group by reflectional symmetry
before rotational symmetry and translational symmetries.

\noindent
{\bf G3 (Proximity in symmetry clique):}
If {\bf G1} and {\bf G2} cannot set a precedence, e.g., between rotational and translational
symmetries of the same order, then grouping of part ensembles closer in proximity takes precedence.

%

\noindent
{\bf A1 (Symmetry preservation):}
Symmetry-preserving assembly takes precedence over symmetry-breaking assembly.

\noindent
{\bf A2 (Connectivity strength):}
If {\bf A1} cannot set a precedence, then order assembly operations according
to a geometric {\em connectivity strength\/} measure.

\end{document}